\documentclass[10pt, journal, final, letterpaper, oneside,
twocolumn]{IEEEtran}

\usepackage{epsfig,latexsym,amsfonts,amsmath,amssymb,verbatim,cite,mathrsfs}
\usepackage{graphicx}
\usepackage[dvips]{color}

\def\rR{{\mathbb R}}
\def\eE{{\mathbb E}}

\def\QED{\mbox{\rule[0pt]{1.5ex}{1.5ex}}}

\def\endproof{\hspace*{\fill}~\QED\par\endtrivlist\unskip}

\def\@begintheorem#1#2{\tmpitemindent\itemindent\topsep 0pt\rm\trivlist
    \item[\hskip \labelsep{\indent\it #1\ #2:}]\itemindent\tmpitemindent}
\def\@opargbegintheorem#1#2#3{\tmpitemindent\itemindent\topsep 0pt\rm \trivlist
    \item[\hskip\labelsep{\indent\it #1\ #2\
    \rm(#3):}]\itemindent\tmpitemindent}
\def\@endtheorem{\endtrivlist\unskip}

\newtheorem{definition}{Definition}[section]
\newtheorem{remark}{Remark}[section]
\newtheorem{theorem}{Theorem}[section]
\newtheorem{fact}[theorem]{Fact}
\newtheorem{proposition}[theorem]{Proposition}
\newtheorem{corollary}[theorem]{Corollary}
\renewcommand{\theequation}{\arabic{section}.\arabic{equation}}
\newcommand{\Section}[1]{\section{#1}
\setcounter{equation}{0}}

\newcommand{\seclbl}[1]{\label{sec:#1}}
\newcommand{\secref}[1]{Section~\ref{sec:#1}}
\newcommand{\eqnlbl}[1]{\label{eq:#1}}
\newcommand{\eqnref}[1]{(\ref{eq:#1})}

\begin{document}

\title{On the existence and characterization of the maxent
distribution under general moment inequality constraints}

\author{Prakash Ishwar,~\IEEEmembership{Member,~IEEE,} and Pierre
Moulin,~\IEEEmembership{Fellow,~IEEE}\thanks{ Manuscript received
November 14, 2001; first revision July 2003; second revision December
2004.  This work was supported by NSF under grants MIP-97-07633 and
CDA-96-24396.}  \thanks{P.~Ishwar was with the Beckman Institute and
the Department of Electrical and Computer Engineering, University of
Illinois, Urbana, IL 61801 USA. He is now in the Department of
Electrical and Computer Engineering, Boston University, Boston, MA
02215 USA (e-mail:~pi@bu.edu).}  \thanks{P.~Moulin is with the Beckman
Institute, Coordinated Science Lab, and the Department of Electrical
and Computer Engineering, University of Illinois, Urbana, IL 61801 USA
(e-mail:~moulin@ifp.uiuc.edu).}}

\date{today}

\maketitle

\thispagestyle{plain}
\pagestyle{plain}

\begin{abstract}
A broad set of sufficient conditions that guarantees the existence of
the maximum entropy (maxent) distribution consistent with specified
bounds on certain generalized moments is derived.  Most results in the
literature are either focused on the minimum cross--entropy
distribution or apply only to distributions with a bounded--volume
support or address only equality constraints. The results of this work
hold for general moment inequality constraints for probability
distributions with possibly unbounded support, and the technical
conditions are explicitly on the underlying generalized moment
functions. An analytical characterization of the maxent distribution
is also derived using results from the theory of constrained
optimization in infinite--dimensional normed linear spaces. Several
auxiliary results of independent interest pertaining to certain
properties of convex coercive functions are also presented.
\end{abstract}

\Section{\seclbl{intro}Introduction} 

Consider the problem of estimating a signal from ``noisy''
observations when we have complete information about the statistics of
the observation process but only partial prior (statistical)
information about the signal of interest.  Partial prior information
about the signal probability distribution might be available in the
form of bounds on a restricted set of certain general moment
measurements.  Incompleteness in the prior information is with regard
to the underlying signal probability distribution that is consistent
with the measurements.  There arises the question of selecting a
distribution from the feasible ones that is noncommittal with respect
to missing information.  The maxent principle provides a selection
mechanism that enjoys several appealing optimality properties
\cite{topsoe79,shore-IT80,campenhout,jaynes-PrIEEE82,csiszar-AnnSt91,vogel,grunwald}.

Questions of existence and characterization of the maxent distribution
in a collection of probability distributions over a
finite--dimensional Euclidean space are, in general, problems in
infinite dimensional constrained optimization involving several
subtleties, and many derivations in the literature contain
errors\footnote{See Borwein and Limber \cite{borwein} for references
to these nonrigorous derivations.}. Although the form of the maxent
distribution subject to general moment {\em equality} constraints has
been known for long, there has been little systematic investigation
into its validity and the existence of the maxent distribution.  Most
results in the literature are either focused on the minimum
cross--entropy distribution or apply only to distributions with a
bounded--volume support. A key difficulty in extending such existence
and characterization results from cross--entropy to differential
entropy is that unlike cross--entropy which is always well-defined,
nonnegative, and satisfies a joint lower semi--continuity property,
differential entropy is not always well--defined and lacks a crucial
upper--semicontinuity property that is needed for establishing
existence results along the lines of those for cross--entropy.

Building upon results due to Csisz\'{a}r and Tops\o{e}
\cite{csiszar-AnnPr75,topsoe79}, we provide broad sufficient
conditions on {\em general convex families} of distributions that
guarantee the {\em existence} of the maxent distribution in the
family. We also specialize these existence results to specific convex
families of probability distributions defined through general moment
inequality constraints.  We also provide an analytical {\em
characterization} of the maxent distribution for such general
moment--constrained families. Our existence and characterization
results hold for probability densities over a finite--dimensional
Euclidean space, that is, finite--dimensional probability
distributions that are absolutely continuous with respect to the
Lebesgue measure, although they can be extended to general
finite--dimensional sigma--finite measures also. For results
pertaining to specific convex families of distributions defined
through general moment inequality constraints, a finite number of
constraints is assumed although the results can be extended when there
are a countable number of constraints. Our results apply for both
differential entropy and I--divergence although we state and prove
results only for differential entropy.

Existence and characterization results for a family of compactly
supported probability densities on the real line with a prescribed
mean and variance (moment equality constraints) are presented in
\cite{DowsonW_ME:IT73}. The analysis in \cite{csiszar-AnnPr75} is
exclusively devoted to I--divergence (which requires a reference
measure) and not differential entropy and the existence results were
stated only in terms of the convexity and variational completeness of
the feasible set of distributions.  Unlike the results in
\cite{csiszar-AnnPr75} which are in terms of general conditions on the
convex collections of distributions satisfying general moment
constraints with {\em equality}, which might be difficult to check in
practice, our results are for general moment {\em inequality}
constraints, and the technical conditions are {\em explicitly} on the
underlying moment functions\footnote{We use the terms moment function
and measurement function interchangeably.}.  The results presented in
\cite{topsoe79} hold for probability distributions over a {\em
countable space} and the existence results therein pertain to the {\em
center of attraction} of a convex collection of distributions. The
relationship between the center of attraction of a family of densities
defined via moment {\em equality} constraints\footnote{The moments
were with respect to a $\sigma$--finite reference measure over a
general measurable space.}  and the maximum--likelihood estimate in an
associated exponential family of densities is derived in
\cite{JuppM_AN:ScandJStat83}.  Borwein and Limber in \cite{borwein}
also provide a set of sufficient conditions for the existence of the
maxent distributions and characterize its form but these results
differ from ours in several aspects. Their results are for equality
constraints, ours are for inequality constraints. The underlying space
in their analysis is the real line, our analysis is on $\rR^d$. Their
analysis considered distributions with bounded support. Our analysis
allows distributions with unbounded support.

For a collection of distributions satisfying moment--equality
constraints, the maxent distribution, when it exists, has an
exponential form where the exponent belongs to the closed subspace
spanned by the measurement functions \cite{borwein}. The additional
flexibility allowed by inequality constraints leads to a stronger
characterization of the maxent distribution. We show, not
surprisingly, that under moment inequality constraints and mild
regularity assumptions, the maxent distribution has an exponential
form where the exponent belongs to the negative cone generated by the
measurement functions. In many applications, inequality constraints
are perhaps more commonly encountered than equality constraints. With
equality constraints, it is often difficult to verify the existence of
a maxent solution because of possible errors in the estimated
moments. The conditions of our existence and characterization theorem
are application--oriented in the sense that if the measurement
functions meet certain general requirements, the maxent solution
exists and has a special exponential form.  We have learnt (thanks to
an anonymous reviewer) about another work by Csisz\'{a}r which
addresses inequality constraints\cite{csiszar-AnnPr84}. However, those
results are for the minimum cross--entropy problem and it is not clear
how they could be extended to the maxent problem especially when the
support set has unbounded volume -- an important consideration in our
work. Other general references where inequality constraints have been
considered include \cite[Section~13.1.4]{Kapur_ME93} and
\cite{Khudanpur_AM95}.

We provide two sets of sufficient conditions on the underlying
constraint functions that guarantee the existence of the maxent
distribution. In one set of sufficient conditions, the proof hinges on
the assumption that the distributions of interest have supports that
are contained in a finite volume subset of $\rR^d$ that need not be
bounded. The second set of sufficient conditions removes this
restriction by assuming the presence of a general ``stabilizing''
moment constraint in the definition of the feasible collection of
distributions. We also present a rich class of ``well--behaved''
functions that provide the general ``stabilizing'' moment constraints
guaranteeing the existence of the maxent distribution.  Frequently
encountered constraints such as mean quadratic energy and mean
absolute energy are well--behaved. These well--behaved constraints
have several interesting and intuitively appealing properties that are
of independent interest.

In \secref{maxentest} we provide some background, define all important
terms, and state the maxent problem. In \secref{mainthm} we state the
main results of this work -- fundamental theorems on the existence and
characterization of the maxent distribution consistent with specified
moment inequality constraints. Proofs of these theorems and related
results of independent interest are presented in the appendices.

\Section{\seclbl{maxentest}Background and problem statement}

{\bf Notation:} $\rR$ denotes the set of real numbers, 
\[
\overline{\rR} := \rR \bigcup \{+\infty,-\infty\}
\]
the set of extended real numbers, and $\rR^d$ the $d$--dimensional
real Euclidean space. Vectors are denoted by boldface letters, for
example, ${\bf x} \in {\rR}^{d}$, and finite dimensional vectors are
treated as column vectors. All sets in this work are
Lebesgue--measurable. If $A$ and $B$ are Lebesgue--measurable subsets
of $\rR^d$, then the statement $A=B$ means that the set of points not
simultaneously in both $A$ and $B$ has Lebesgue measure zero and $A$
is said to be equal to $B$ almost everywhere (a.e.). All functions in
this work take values in $\overline{\rR}$ and are measurable with
respect to the Lebesgue measure over ${\rR}^{d}$. Inequalities
involving measurable functions are to be understood in the a.e. sense.
All integrals are in the sense of Lebesgue. A probability density
function (pdf) is a measurable function $\pi({\bf x})$ on ${\rR}^{d}$
that is non--negative almost everywhere (a.e.) and integrates to unity
over ${\rR}^{d}$.  All results in this work are stated for probability
densities over finite--dimensional Euclidean spaces, that is,
probability distributions that are absolutely continuous with respect
to the Lebesgue measure, although they can be extended to general
sigma--finite measures on $\rR^d$ also. ${\mathcal{L}}^{1}({\rR}^{d})$
and $\mathcal{L}^{\infty}(\rR^d)$ respectively denote the set of
absolutely--integrable functions over ${\rR}^{d}$ and the set of
essentially bounded functions \cite[p.~119]{royden} over $\rR^d$. For
convenience, we shall often omit the `${\bf x}$' and the
`$\mathrm{d}{\bf x}$' that appear inside an integral. Thus,
\[
\int_{A}f({\bf x})\mathrm{d}{\bf x}
\]
will often be abbreviated to $\int_A f({\bf x})$ or simply $\int_A
f$. The symbol $\pi$ and its variants will denote pdfs and
\[
\eE_{\pi}[\phi] := \int_{\rR^d} \phi\cdot\pi
\]
denotes the mathematical expectation of the function $\phi({\bf x})$
under the pdf $\pi({\bf x})$. The support of a function $f({\bf x})$
is the set of points where it is nonzero\footnote{Note that we are
working with probability density functions.} and is denoted by
$\mathrm{supp}(f)$. The indicator or characteristic function of a
subset $A$ of $\rR^d$ denoted by ${\bf 1}_{A}({\bf x})$ is the
function that is equal to one over $A$ and zero elsewhere. The volume
of a Lebesgue--measurable subset $S$ of $\rR^d$ is its Lebesgue
measure and is denoted by $|S|$. In addition to the arithmetic of the
extended reals, the following conventions regarding infinity are
adopted in keeping with measure--theoretically consistent operations:
\begin{equation}
\begin{array}{ccc}
\ln 0 = -\infty, & \ln\frac{a}{0} = +\infty,\,\forall a > 0, &
0\cdot(\pm\infty) = 0.\nonumber
\end{array}
\end{equation}
Thus $0\ln 0 = 0$ which also agrees with the limiting value of the
quantity $t\ln t$ as the variable $t$ decreases to zero.

In Bayesian inference, signals of interest are modeled as
high--dimensional real random vectors with associated pdfs referred to
as prior distributions on the signals.  Let ${\bf X}\,\in\,\rR^d$ have
an underlying $d$--dimensional pdf denoted by ${\pi}({\bf x})$.  In
many applications, only limited information about $\pi({\bf x})$ can
be gathered.  Moments of probability distributions are often used to
describe the underlying statistical structure of a stochastic
process. For example, the set of all finite--order moments of a scalar
random variable provides, under suitable regularity assumptions, a
complete statistical description of the random variable
\cite[Theorem~30.1, p.~388]{billey}.  In practice, only a finite set
of moments is a priori known or can be estimated (measured) from
samples. In many cases even these are not available but bounds on the
moments are available. The bounds may be regarded as arising from the
impreciseness of moment measurements.  For example, for $p>0$, the
empirical mean $\ell^p$ energies of wavelet coefficients in different
subbands are often used to construct statistical models for images
\cite{icip99,icassp00,icip00}. In general, the limited information
will be unable to single out a desirable distribution that is
consistent with the moment constraints. The limited information would
rather specify a whole class of distributions that satisfy the moment
constraints.

Let prior information about a random vector {\bf X} be available in
terms of upper bounds on the expected values of certain real--valued
Lebesgue--measurable (measurement) functions
\[
\phi_{\gamma}:\rR^d\rightarrow \rR,\,\gamma\in\Gamma,
\]
where $\Gamma$ is a finite index set\footnote{The focus of this work
is on the case when the number of measurement functions is finite but
the results can also be extended to the case when there are a
countable number of measurement functions.}. A useful notion is that
we can sometimes design these functions $\phi_\gamma({\bf x})$ (that
is, the measurements). Each candidate distribution $\pi({\bf x})$ that
is consistent with these measurements then belongs to the set
\begin{eqnarray}
\Omega({\bf u}) &:=& \{\mathrm{pdf}\ \pi: \mathrm{supp}(\pi) \subseteq
S,\ \mbox{and for all $\gamma$ in $\Gamma$},\nonumber\\
&&\mbox{}\eE_{\pi}[\phi_{\gamma}] \leq u_{\gamma} < +\infty\},
\eqnlbl{omegadef}
\end{eqnarray}
where $S$ is a closed Lebesgue--measurable subset of $\rR^d$ having
nonzero but possibly infinite volume and 
\[
{\bf u}:= \{u_\gamma\in\rR\}_{\gamma\in\Gamma} 
\]
is a finite--dimensional, real--valued, vector of moment upperbounds.
We assume that the only prior information available is expressed by
the moment constraints of $\Omega$. Since $\Omega$ is defined through
inequality constraints that are linear in $\pi$, {\em it is a convex
set of probability distributions}. It is possible to implicitly
incorporate support constraints into $\Omega$ through appropriate
moment inequalities without explicitly requiring that
$\mathrm{supp}(\pi) \subseteq S$ in the definition. For example, if
\[
u_0 = u_1 = 0,
\]
and
\[
\phi_{0}({\bf x}) := -\phi_{1}({\bf x}) := 1 - {\bf 1}_{S}({\bf x})
\]
then for each $\pi$ belonging to $\Omega$, we have $\left|
\mathrm{supp}(\pi) \backslash S \right| = 0$.  For clarity of
exposition we shall primarily work with the convex collection
\eqnref{omegadef}. However, it is quite straightforward to extend our
results to convex collections having individual lowerbounds
$\{l_\gamma\in\rR\}_{\gamma\in\Gamma}$ on the moment measurements.

In general, many distributions will satisfy the moment constraints of
$\Omega$.  The choice of a distribution from this moment consistent
class depends upon the goals to be achieved by the selection. For the
application of lossless compression, a clear answer can be given.  The
unique pdf that maximizes the differential entropy functional 
\[
h(\pi) := -\eE_\pi[\ln\pi]
\]
over a convex set $F$, {\em whenever it exists}, also minimizes the
worst--case rate for encoding repeated independent observations of
${\bf X}$ ``losslessly''
\cite[pp.~105--106]{my_phd_thesis},\cite[pp.~61--63]{grunwald},\cite[Theorem~3,
p.~16]{topsoe79} (The results in \cite{grunwald,topsoe79} are for
discrete entropy).  A similar result holds for high--rate lossy
compression \cite{vogel}.

\begin{definition}{(Maximum entropy distribution)} {\rm
Let $F$ be a convex collection of distributions for which
\[
F\cap\{\mathrm{pdf}\ \pi: h(\pi) > -\infty\}
\]
is nonempty.  The maxent distribution in $F$ whenever it exists is the
unique pdf $\pi_{ME}$ belonging to $F$ satisfying\footnote{The
subscript ME stands for maximum entropy.}
\[
h(\pi_{ME}) = \max_{\pi\in F}h(\pi).
\]}
\end{definition}

It may be noted that since $h(\pi)$ is a concave functional
\cite{cover-InfThyJWly}, the set 
\[
\{\mathrm{pdf}\ \pi: h(\pi) > -\infty\}
\]
is convex.  The uniqueness of $\pi_{ME}$ follows from the strict
concavity of the differential--entropy functional
\cite{cover-InfThyJWly} and the convexity of $F$.

In addition to being minimax optimal for the application of lossless
compression with uncertain source statistics discussed above, the
maxent distribution is also ``maximally noncommittal'' with respect to
missing information while satisfying prior constraints
\cite{jaynes-PrIEEE82}.  Shore and Johnson in \cite{shore-IT80} show
that if a distribution has to be picked from a class of probability
distributions by maximizing a functional satisfying some natural
postulates, it must necessarily be the maxent functional.  Again, in a
study of logically consistent methods of inference, Csisz\'{a}r
demonstrates that the maxent distribution is the only one that
satisfies two different intuitively appealing axiom systems
\cite{csiszar-AnnSt91}. These properties of the maxent distribution
make it a desirable choice for signal estimation.

In some applications, based on previous measurements, a reliable
reference distribution $r({\bf x})$ for the signal of interest is
available.  New moment measurements might reveal that the reference
distribution has inconsistencies with new information in the form of
bounds on moments \eqnref{omegadef}. The situation suggests a revision
of the reference model while not ignoring earlier measurements. An
attractive model selection criterion in this situation is to select
the distribution in $\Omega$ that is closest to the reference
distribution in the sense that it has minimum cross--entropy (MCE)
relative to the reference prior:

\begin{definition}{(Cross--entropy \cite[p.~146]{csiszar-AnnPr75})} {\rm
The {\em cross--entropy} of pdf $\pi_1({\bf x})$ with respect to pdf
$\pi_2({\bf x})$ (also known as the I--divergence, Kullback--Leibler
distance, relative entropy, and information discrimination) denoted by
$D(\pi_{1}||\pi_{2})$ is defined as:
\begin{equation}
D(\pi_1||\pi_2) := \left\{\begin{array}{ll}
                          \eE_{\pi_{1}}[\ln(\frac{\pi_{1}}{\pi_{2}})] &
                          \mbox{if $\pi_1 \ll \pi_2$
                          (see~Definition~\ref{def:abscont})} \\
                          +\infty  & \mbox{otherwise.}
                         \end{array}
                  \right.
\nonumber
\end{equation}
}
\end{definition}

\begin{definition}{(I--projection \cite[p.~147]{csiszar-AnnPr75})} {\rm
Let $r$ be a pdf and $F$ a convex collection of priors such that
\[
F\cap\{\mathrm{pdf}\ \pi: D(\pi||r) < +\infty\}
\]
is nonempty.  The I--projection of $r$ onto $F$, whenever it exists,
is the unique pdf $\pi_{MCE}$ belonging to $F$ satisfying
\[
D(\pi_{MCE}||r) = \min_{\pi\in F}D(\pi||r).
\]
}
\end{definition}

The updated distribution $\pi_{MCE}$ is referred to as the {\em
I--projection} of $r$ onto $F$.  Since $D(\pi||r)$ is strictly convex
in $\pi$ \cite{cover-InfThyJWly}, and $F$ is a convex set, $\pi_{MCE}$
is unique whenever it exists.

Generally speaking, the maxent distribution in $\Omega$
\eqnref{omegadef} need not exist. Our goal is to provide a set of
sufficient conditions on the measurement functions that guarantee the
existence of the maxent prior. We provide such a set of conditions in
the following section.  We also characterize the form of the maxent
prior. Similar existence and characterization results for
I--projection under moment {\em inequality} constraints can be derived
along similar lines but are omitted from the present work
(see \cite{csiszar-AnnPr75,csiszar-AnnPr84,CsiszarM_IP:IT03,CsiszarM_OI:IT04}).

\Section{\seclbl{mainthm}Existence and characterization of the maxent
distribution}

The following theorem proved in Appendix~B.1 provides a
characterization of the unique maxent distribution in $\Omega$ subject
to suitable technical conditions.

\begin{theorem}{(Characterization of the maxent distribution)}\label{theorem:charME}
Let $\Omega({\bf u})$ be as in \eqnref{omegadef}. Let there exist a
pdf $\pi_0$ in $\Omega({\bf u})$ such that for all $\gamma$ in
$\Gamma$, $\eE_{\pi_0}\left[\phi_\gamma\right] < u_\gamma$. If the
unique maxent pdf $\pi_{ME}$ belonging to $\Omega({\bf u})$ exists and
$h(\pi_{ME})$ is finite, then the maxent pdf has the form
\begin{eqnarray}
\lefteqn{\pi_{ME}({\bf x},{\bf u}) = {\bf 1}_{S_{ME}}({\bf x})\cdot }
\nonumber\\ & & \cdot \exp\left\{-\alpha({\bf u}) -
\sum_{\gamma\in\Gamma}\lambda_\gamma({\bf u})\phi_{\gamma}({\bf
x})\right\}, \eqnlbl{MEprior}
\end{eqnarray}
where $S_{ME} := \mathrm{supp}(\pi_{ME}) \subseteq S$ satisfies
$\eE_{\pi}\left[{\bf 1}_{S\backslash S_{ME}}\right] = 0$ for every
$\pi\in\Omega({\bf u})$ for which $-\infty < h(\pi)$ and 
\[
\alpha({\bf u}) = \ln\left( \int_{S_{ME}}
\exp\left\{-\sum_{\gamma\in\Gamma}\lambda_\gamma({\bf
u})\phi_{\gamma}({\bf x})\right\}\mathrm{d}{\bf x}\right)
\]
is a finite normalization constant. The parameters
$\{\lambda_\gamma({\bf u})\}_{\gamma\in\Gamma}$ are all nonnegative,
and satisfy
\begin{eqnarray}
\eqnlbl{qualif}
\sum_{\gamma\in\Gamma}\lambda_\gamma(\eE_{\pi_{ME}}\left[\phi_\gamma\right] -
u_\gamma) = 0.
\end{eqnarray}
Moreover, 
\begin{eqnarray*}
h(\pi_{ME}) &=& \alpha({\bf u}) +
\sum_{\gamma\in\Gamma}\lambda_\gamma( {\bf u})
\eE_{\pi_{ME}}\left[\phi_\gamma\right] \\ &=& \alpha({\bf u}) +
\sum_{\gamma\in\Gamma}u_\gamma \lambda_\gamma({\bf u}).
\end{eqnarray*}
\end{theorem}

\begin{remark}\label{remarks:suppremark} {\rm
Note that if $\pi$ belongs to $\Omega({\bf u})$ and $-\infty <
h(\pi)$, then $\pi \ll \pi_{ME}$.  If there exists a pdf $\pi$ in
$\Omega({\bf u})$ with $-\infty < h(\pi)$ and $\mathrm{supp}(\pi) =
S$, then the set $S\backslash S_{ME}$ has zero volume; that is,
$S_{ME}$ almost everywhere coincides with $S$ and we may take $S_{ME}
= S$ in the above theorem.}
\end{remark}

\begin{remark} {\rm 
The numbers $\{\lambda_\gamma\}_{\gamma\in\Gamma}$ in
Theorem~\ref{theorem:charME} above are Lagrange multipliers associated
with the moment constraints of $\Omega({\bf u})$ in \eqnref{omegadef}.
The constraint qualification \eqnref{qualif} implies that
$\lambda_\gamma = 0$ if constraint $\gamma$ is inactive, that is,
$\eE_{\pi_{ME}}\left[\phi_\gamma\right] < u_\gamma$.}
\end{remark}

\begin{remark} {\rm 
Since $S$ has nonzero volume and $\pi_{ME}$ is unique, if the
measurement functions $\{\phi_\gamma\}_{\gamma\in\Gamma}$ are linearly
independent then there is a unique choice for the parameters
$\boldsymbol{\lambda} := \{\lambda_\gamma\}_{\gamma\in\Gamma}$ that
satisfies the moment constraints of $\Omega({\bf u})$. In this case,
the mapping from the vector of moment bounds ${\bf u}$ to the vector
of Lagrange multipliers $\boldsymbol{\lambda}$ is a function, that is,
it is not a one-to-many map. If the measurement functions are not
linearly independent, the characterization theorem still holds, but
the Lagrange multipliers need not be unique.}
\end{remark}

\begin{remark}\label{remarks:basicprops} {\rm 
The Lagrange multipliers $\boldsymbol{\lambda}({\bf u})$ are usually
implicit functions of the moment bounds ${\bf u}$.  If for some value
of ${\bf u}$ a Lagrange multiplier turns out to be zero --- that is,
$\lambda_\gamma({\bf u}) = 0$ for some $\gamma\in\Gamma$ (a situation
that will arise if the associated moment constraint is inactive, that
is, $\eE_{\pi_{ME}}\left[\phi_\gamma\right] < u_\gamma$) --- then the
maxent solution corresponding to any larger value of $u_\gamma$ will
remain the same (see Appendix~B.2 for a proof).  Thus, the map
$\boldsymbol{\lambda}({\bf u})$ from moment bounds to Lagrange
multipliers is in general not injective. However, see the following
remark.}
\end{remark}

\begin{remark} {\rm 
The mapping from the moment upper--bounds ${\bf u}$ to the Lagrange
multipliers $\boldsymbol{\lambda}({\bf u})$ is one-to-one when the
domain is restricted to the set of those values of ${\bf u}$ for which
$\lambda_\gamma({\bf u}) > 0$ for every $\gamma$ in $\Gamma$, that is,
{\em all} the constraints are active. This fact can be seen by the
following argument. Suppose that
$\{u^{(1)}_\gamma\}_{\gamma\in\Gamma}$ and
$\{u^{(2)}_\gamma\}_{\gamma\in\Gamma}$ both map to the same set of
strictly positive Lagrange multipliers $\{\lambda_\gamma >
0\}_{\gamma\in\Gamma}$. Then because all constraints are active, due
to \eqnref{qualif}, necessarily 
\[
u^{(1)}_\gamma = \eE_{\pi_{ME}}\left[\phi_\gamma({\bf x})\right] =
u^{(2)}_\gamma
\]
for every $\gamma$ in $\Gamma$.}
\end{remark}

Theorem~\ref{theorem:charME} asserts that whenever the maxent
distribution in a moment--consistent class exists then, subject to
some mild technical conditions, it has a natural exponential form
given by \eqnref{MEprior}. The next result proved in Appendix~B.3
essentially asserts that if a pdf having the exponential form given by
\eqnref{MEprior} is moment consistent then it must be the maxent
distribution for the moment--consistent class. In this sense, the next
result is a converse to Theorem~\ref{theorem:charME}.
\begin{theorem}{(Converse to the characterization theorem)}\label{theorem:converse}
Let $\Omega({\bf u})$ be as in \eqnref{omegadef}. Consider a pdf
\begin{eqnarray}
\pi_{\exp}({\bf x},\boldsymbol{\lambda}) &:=& {\bf 1}_{S_{\exp}}({\bf
x}) \cdot\exp\left\{-\alpha -\sum_{\gamma\in\Gamma}\lambda_\gamma({\bf
u})\phi_{\gamma}({\bf x})\right\}, \nonumber
\end{eqnarray}
where $S_{\exp}$ is a measurable subset of $S$, and the vector of
nonnegative but finite--valued parameters $\{\lambda_\gamma({\bf
u})\}_{\gamma\in\Gamma}$ is denoted by $\boldsymbol{\lambda}$.  If 
\begin{itemize}
\item[(i)] $\pi_{\exp}$ belongs to $\Omega({\bf u})$, 
\item[(ii)] $\eE_{\pi}\left[{\bf 1}_{S\backslash S_{\exp}}\right] = 0$
for every $\pi\in\Omega({\bf u})$ for which $-\infty < h(\pi)$, and
\item[(iii)]
\begin{eqnarray}
\sum_{\gamma\in\Gamma}\lambda_\gamma(\eE_{\pi_{\exp}}\left[\phi_\gamma\right] -
u_\gamma) = 0, \nonumber
\end{eqnarray}
\end{itemize}
then $\pi_{\exp}$ is the unique maxent pdf in $\Omega({\bf u})$ and
\[
h(\pi_{\exp}) = \alpha + \sum_{\gamma\in\Gamma}\lambda_\gamma({\bf
u})\eE_{\pi_{\exp}}[\phi_\gamma]
\]
is finite.
\end{theorem}

Before entering into sufficient conditions for the existence of the
maxent distribution, we would like to briefly comment on some
practical aspects of computing the Lagrange multipliers from given
moment constraints. The infinite--dimensional constrained entropy
maximization problem can be converted to a finite--dimensional convex
minimization problem by invoking Lagrange duality theory
\cite[pp.~21--24]{my_phd_thesis}. This forms the basis for developing
numerical techniques for computing the Lagrange multipliers that
characterize the maxent distribution. Several algorithms based on
iterative gradient--projection or moment--matching procedures having
different convergence properties have been proposed in the literature,
for example, Bregman's balancing method, multiplicative algebraic
reconstruction technique, generalized iterative scaling method,
Newton's method, interior--point methods,
etc. \cite{entoptmathprog}. However, these algorithms have been
largely applied to problems where the underlying space is a finite set
and require evaluating moments at each step. This task can be
nontrivial if the underlying space is $\rR^d$ and $d$ is large, as in
the case of images, because moment computation will involve evaluating
very high dimensional integrals. One would typically need to take
recourse to computationally intensive algorithms like importance
sampling or Markov--chain Monte--Carlo for numerically evaluating the
high--dimensional integrals at each step. However, in certain
situations it might be possible to take advantage of the structure of
the specific moment functions to develop fast heuristic approximations
for the Lagrange multipliers \cite[Chapter~4]{my_phd_thesis},
\cite{icip99,icassp00,icip00}.

\begin{theorem}{(Existence of the maxent distribution -- finite volume
support constraint)}\label{theorem:existence} Let $S$ be a closed,
Lebesgue--measurable subset of $\rR^d$ having nonzero but finite
volume.  If $F$ is a nonempty, convex, $\mathcal{L}^1$--complete
collection of pdfs over $S$ and $-\infty < h(\pi_0)$ for at least one
pdf $\pi_0$ belonging to $F$, then 
\[
h(F) := \sup_{\pi\in F}h(\pi) \in \rR,
\]
that is, $h(F)$ is finite, and there exists a unique maxent pdf in
$F$.
\end{theorem}

\begin{corollary}\label{corollary:existence}
Let $\Omega({\bf u})$ be as in \eqnref{omegadef}. Let
$\{\phi_\gamma\}_{\gamma\in\Gamma}$ be uniformly bounded from below by
$L\in\rR$ and $S$ have nonzero but finite volume.  If $\Omega$ is
nonempty and
\[
C:= \bigcap_{\gamma\in\Gamma} \{{\bf x}\in S: \phi_\gamma({\bf x})\leq
u_\gamma \}
\]
has nonzero volume then there exists a unique maxent pdf $\pi_{ME}$ in
$\Omega({\bf u})$ having the exponential form given by
Theorem~\ref{theorem:charME} with $h(\pi_{ME}) \in \rR$.
\end{corollary}

The proof of Theorem~\ref{theorem:existence} appears in Appendix~C.1.
The proof of Corollary~\ref{corollary:existence} appears in
Appendix~C.2.  While the finite measure condition is crucial to the
proof of Theorem~\ref{theorem:existence} and
Corollary~\ref{corollary:existence}, the next theorem and corollary
show that the existence of the maxent distribution is guaranteed by
the presence of a ``stabilizing'' constraint function in the
definition of $\Omega$ even if the support set's volume is not
finite. The proofs of these results appear in Appendix~C.3 and
Appendix~C.4 respectively.  We would like to point out that the
sufficient conditions for existence mentioned in
\cite{csiszar-AnnPr75} and the corollary following Theorem~5.2 in
\cite{csiszar-bayes} for the cross-entropy problem is not available
for differential entropy unless attention is restricted to
distributions supported on a set of finite Lebesgue measure due to the
lack of a general upper-semicontinuity property for differential
entropy. It is not immediately clear how those results can be extended
to distributions having an infinite-volume support.

\begin{definition}{(Stable function)} {\rm 
A real--valued measurable function $f({\bf x})$ is stable if
$\exp\{-\lambda f({\bf x})\}$ belongs to $\mathcal{L}^1(\rR^d)$ for
all $\lambda > 0$.}
\end{definition}

\begin{remark} {\rm 
If $f({\bf x})$ is stable so is $\lambda f({\bf x})$ for all $\lambda
\in(0,+\infty)$.}
\end{remark}

\begin{theorem}{(Existence of the maxent distribution -- stabilizing 
constraint)}\label{theorem:existence-2} Let $S$ be a closed,
Lebesgue--measurable subset of $\rR^d$ having nonzero but possibly
infinite volume and $F$ be a nonempty, convex,
$\mathcal{L}^1$--complete collection of pdfs over $S$.  If 
\begin{itemize}
\item[(i)] there exists a $\pi_0$ in $F$ such that $ -\infty <
h(\pi_0)$ and 
\item[(ii)] there exist finite reals $L$, $u$, with $L \leq u$, and a
stable function $\psi$ such that for all $\pi$ in $F$, $L \leq
\eE_\pi[\psi] \leq u$, 
\end{itemize}
then $h(F) := (\sup_{\pi\in F}h(\pi)) \in \rR$, that is, $h(F)$ is
finite, and there exists a unique maxent pdf in $F$.
\end{theorem}

\begin{corollary}\label{corollary:existence-2} 
Let $\Omega({\bf u})$ be as in \eqnref{omegadef}. Let
$\{\phi_\gamma\}_{\gamma\in\Gamma}$ be uniformly bounded from below by
$L\in\rR$ and $S$ have nonzero (but possibly infinite) volume.  If
$\Omega$ is nonempty and
\begin{enumerate}
  \item $C:= \cap_{\gamma\in\Gamma} \{{\bf x} \in S: \phi_\gamma({\bf x})\leq u_\gamma \}$ has nonzero volume,
  \item there exists $\gamma_0 \in\Gamma$ for which $u_{\gamma_0} \in
       \rR$ and $\phi_{\gamma_0}$ is stable,
\end{enumerate}
then there exists a unique maxent pdf $\pi_{ME}\in\Omega({\bf u})$
having the exponential form given by Theorem~\ref{theorem:charME} with
$h(\pi_{ME}) \in \rR$.
\end{corollary}

\begin{remark} {\rm
In Corollaries~\ref{corollary:existence} and
\ref{corollary:existence-2}, the condition that the measurement
functions $\{\phi_\gamma\}_{\gamma\in\Gamma}$ be uniformly bounded
from below by $L\in\rR$ is sufficient to ensure that $\Omega({\bf u})$
is complete under the $\mathcal{L}^1(\rR^d)$ norm (see
Proposition~\ref{proposition:closedness} in
Appendix~\ref{appendix:existproofs}). The condition that 
\[
C:= \cap_{\gamma\in\Gamma} \{{\bf x}\in S: \phi_\gamma({\bf x})\leq
u_\gamma \}
\]
has nonzero volume is a sufficient condition to ensure that there is
at least one pdf $\pi_0$ with $-\infty < h(\pi_0)$.}
\end{remark}

In conclusion, we demonstrate a rich class of ``well--behaved''
constraint functions for which condition (2) in
Corollary~\ref{corollary:existence-2} is satisfied. The main result
here is Theorem~\ref{theorem:wellbehavedthm} whose proof appears in
Appendix~\ref{appendix:wellbehaved}.

\begin{definition}{(Omni--directional unboundedness)} {\rm 
A real--valued function on a vector space is {\em asymptotically
positive and unbounded in all directions} if $f({\bf z}) \rightarrow
+\infty$ whenever $||{\bf z}|| \rightarrow \infty$. For simplicity we
shall refer to this as the {\em omnidirectional unboundedness}
property (which is also sometimes referred to as the coercive property
\cite[Definition~A.4(c), p.~653]{bertsekas-NlinPrgAthSc}).}
\end{definition}

\begin{remark} {\rm 
In a finite--dimensional Banach space such as $\rR^{d}$, all norms are
equivalent \cite[Theorem~23.6, p.~177]{aliprantis}. In other words, if
$||\cdot||_{\mathrm{a}}$ and $||\cdot||_{\mathrm{b}}$ are two norms,
there are positive constants $L > 0$ and $U > 0$ such that 
\[
L||{\bf x}||_{\mathrm{a}} \leq ||{\bf x}||_{\mathrm{b}} \leq U||{\bf
x}||_{\mathrm{a}}
\]
for all ${\bf x}$ in the finite--dimensional Banach space.  Thus in
${\rR}^{d}$,
\[
||{\bf x}||_{\mathrm{a}} \rightarrow \infty \iff {\bf
||x}||_{\mathrm{b}} \rightarrow \infty.
\]
The definition of omnidirectional unboundedness therefore does not
depend upon the specific norm used when the underlying space is finite
dimensional.}
\end{remark}

\begin{definition}{(Well--behaved function)}\label{def:wellbh} {\rm
Let $\phi:\rR^d \longrightarrow \rR$ be a convex and
omni--directionally unbounded function. A real--valued function
$\psi:\rR^d \longrightarrow \rR$ is well--behaved if there exists a
nonnegative real number M such that
\[
\begin{array}{cccc}
\phi({\bf x}) & \leq & \psi({\bf x}), & \forall\,{\bf
x}\in\rR^d:\,||{\bf x}||> M\ \mbox{and}\\ \sup_{||{\bf x}||\leq
M}|\psi({\bf x})| & < & +\infty.&
\end{array}
\]
}
\end{definition}

\begin{remark} {\rm 
A convex and omni--directionally unbounded function is well--behaved.
If $f({\bf x})$ is well--behaved so is $\lambda f({\bf x})$ for all
$\lambda$ belonging to the open interval $(0,+\infty)$.}
\end{remark}

\begin{theorem}\label{theorem:wellbehavedthm}
A well--behaved function is stable. If $\phi_{\gamma_0}$ is
well--behaved and 
\[
\eE_\pi[\phi_{\gamma_0}] \leq u_{\gamma_0} < +\infty
\]
then $h(\pi)$ exists and
\[
h(\pi) \leq u_{\gamma_0} + \ln||e^{-\phi_{\gamma_0}}||_{\mathcal{L}^1}
< \infty.
\]
Hence, if $\phi_{\gamma_0}$ belongs to
$\{\phi_\gamma\}_{\gamma\in\Gamma}$ in
Corollary~\ref{corollary:existence-2} then 
\[
\sup_{\pi\in\Omega({\bf u})}h(\pi) \leq u_{\gamma_0} +
\ln||e^{-\phi_{\gamma_0}}||_{\mathcal{L}^1} < +\infty.
\]
\end{theorem}

\begin{remark}\label{remarks:wellbhextn} {\rm
Suppose that in Corollary~\ref{corollary:existence-2}, none of the
measurement functions $\{\phi_\gamma\}_{\gamma\in\Gamma}$ is
well--behaved, but some nonnegative linear combination of the
measurement functions
\[
\phi_{\boldsymbol{\mu}} := \sum_{\gamma \in \Gamma}
\mu_{\gamma}\phi_\gamma,\ \mbox{where}\ 0 \leq \mu_\gamma < + \infty\
\mbox{for all}\ \gamma\in\Gamma,
\]
is well--behaved. Let $u_{\boldsymbol{\mu}} := \sum_{\gamma\in\Gamma}
\mu_\gamma u_\gamma$ and
\[
\Omega_{\boldsymbol{\mu}} := \{\mathrm{pdf}\ \pi: \leq
\eE_{\pi}\left[\phi_{\boldsymbol{\mu}}\right] \leq
u_{\boldsymbol{\mu}}\}.
\]
It is clear that $\Omega({\bf u}) \subseteq
\Omega_{\boldsymbol{\mu}}$.  Hence, the well--behaved function
$\phi_{\boldsymbol{\mu}}$ and the associated moment constraint
\[
-\infty < L \leq \eE_{\pi}\left[\phi_{\boldsymbol{\mu}}\right] \leq
u_{\boldsymbol{\mu}}
\]
can be included in the set of available moment measurements without
affecting the maxent solution. Although this new constraint is
redundant, it tells us that Theorem~\ref{theorem:wellbehavedthm} can
be applied and the maxent distribution in $\Omega$ exists under the
mild requirements of Corollary~\ref{corollary:existence-2}.}
\end{remark}

\section*{Acknowledgment}
The authors would like to thank Prof.~Tamer Ba\c{s}ar for helpful
discussions regarding optimization in infinite--dimensional spaces,
Prof.~Imre Csisz\'{a}r for clarifying maxent existence results related
to differential entropy in \cite{csiszar-bayes}, Dr.~Raman
Venkataramani for discussions related to omni--directionally unbounded
functions, and the anonymous reviewer for pointing out that our
results do not require the measurement functions to be nonnegative.

\useRomanappendicesfalse
\appendices
\renewcommand{\theequation}{\thesection.\arabic{equation}}
\setcounter{equation}{0}

\renewcommand{\thedefinition}{\Alph{section}.\arabic{definition}}
\renewcommand{\theremark}{\Alph{section}.\arabic{remark}}
\renewcommand{\thetheorem}{\Alph{section}.\arabic{theorem}}

\Section{Preliminaries}

\begin{definition}{(Convex set)} {\rm
A subset $C$ of a vector space is said to be {\em convex} if whenever
${\bf z}_{1}$ and ${\bf z}_{2}$ are in $C$, so is $\alpha{\bf z}_{1} +
(1-\alpha){\bf z}_{2}$ for every $\alpha$ in the closed interval
$[0,1]$.}
\end{definition}

\begin{definition}{(Convex function)} {\rm 
Let $V$ be a vector space.  A functional $f: V \longrightarrow
\overline{\rR}$ is said to be convex if for every $\alpha\in [0,1]$,
and for any ${\bf z}_{1}$ and ${\bf z}_{2}$ belonging to $V$,
\[
f(\alpha{\bf z}_{1} + (1-\alpha){\bf z}_{2}) \leq \alpha f({\bf
z}_{1}) + (1-\alpha)f({\bf z}_{2}).
\]
If equality holds only when ${\bf z}_{1} = {\bf z}_{2}$ then $f$ is
said to be {\em strictly convex}.  If $-f$ is (strictly) convex then
$f$ is said to be (strictly) {\em concave}.}
\end{definition}

\begin{definition}{(Absolute continuity)}\label{def:abscont} {\rm
A pdf $\pi_1$ is said to be absolutely continuous relative to a pdf
$\pi_2$, in symbols $\pi_1 \ll \pi_2$ or $\pi_2 \gg \pi_1$, if for
every Lebesgue measurable subset $A$ of $\rR^d$, $\int_A \pi_2 = 0$
implies $\int_A \pi_1 = 0$ and hence $\mathrm{supp}(\pi_1) \subseteq
\mathrm{supp}(\pi_2)$.}
\end{definition}

\begin{fact}{\cite[p.~5]{Kullback:ITandStat}}
\label{fact:divnonneg}
The cross--entropy of pdf $\pi_1$ relative to pdf $\pi_2$ is always
well defined and non--negative (it could be $+\infty$). The
cross--entropy is zero if and only if $\pi_1 = \pi_2$ almost
everywhere.
\end{fact}

\begin{fact}{\cite{cover-InfThyJWly}}
Differential entropy $h(\pi)$ is strictly concave in
$\pi$. Cross--entropy $D(\pi_1||\pi_2)$ is convex in the pair
$(\pi_1,\pi_2)$ and strictly convex in $\pi_1$.
\end{fact}

\begin{fact}{(Joint lower semi--continuity of cross--entropy
\cite[Section~2.4, Assertion~5]{pinsker})} \label{fact:semicont} If
the pdfs $p_n$ and $q_n$ converge in $\mathcal{L}^1(\rR^d)$ norm to
pdfs $p$ and $q$ respectively as $n\longrightarrow\infty$, then
\begin{eqnarray}
\eqnlbl{semicont}
D(p||q) &\leq& \liminf_{n\rightarrow\infty}D(p_n||q_n).
\end{eqnarray}
\end{fact}

\begin{fact}\cite[p.88]{my_phd_thesis},\cite{JuppM_AN:ScandJStat83},\cite[Theorem~1, p.~14]{topsoe79}:\label{fact:centerofattraction}
If $F\subseteq \mathcal{L}^1(\rR^d)$ is a complete, convex collection
of pdfs and $h(F) := \sup_{\pi\in F}h(\pi)$ is finite, then there
exists a unique distribution $\pi^*$ belonging to $F$ such that for
every sequence $\{\pi_n\}\subseteq F$ for which $h(\pi_n)\rightarrow
h(F)$, we have $\pi_n \rightarrow \pi^*$ in $\mathcal{L}^1(\rR^d)$
norm.
\end{fact}

\begin{fact}{(A fundamental theorem of convex optimization
    \cite[adapted from Theorem~1, p.~217]{luenberger})}
\label{fact:luenberger}
Let $V$ be a vector space and $F$ a convex subset of $V$. Let $f: F
\longrightarrow \overline{\rR}$ be a convex functional on $F$ and
$\{g_\gamma\}_{\gamma\in\Gamma}$ a finite collection of convex
mappings from $F$ into $\overline{\rR}$. Suppose that there exists a
point ${\bf v}_0$ in $F$ such that for all $\gamma\in\Gamma$,
$g_\gamma({\bf v}_0) < 0$ and
\begin{eqnarray}
\eqnlbl{prob1} 
m_0 := \inf_{{\bf v}\in G} f({\bf v})
\end{eqnarray}
is finite where 
\[
G := \{{\bf v}\in F: g_\gamma({\bf v}) \leq
0,\,\forall\gamma\in\Gamma\}.
\]
Then there exist nonnegative Lagrange multipliers
$\{\lambda_\gamma\}_{\gamma\in\Gamma}$ such that
\begin{eqnarray}
\eqnlbl{prob2} 
m_0 = \inf_{{\bf v}\in F}\{f({\bf v}) +
\sum_{\gamma\in\Gamma}\lambda_\gamma g_\gamma({\bf v}) \}.
\end{eqnarray}
Furthermore, if the infimum is achieved in \eqnref{prob1} by ${\bf
v}^{*}$ belonging to $G$, it is also achieved by ${\bf v}^{*}$ in
\eqnref{prob2} and
\begin{eqnarray}
\eqnlbl{kkt}
\sum_{\gamma\in\Gamma} \lambda_\gamma g_\gamma({\bf v}^{*}) = 0.
\end{eqnarray}
\end{fact}

\Section{Characterization of the maxent distribution}

\setcounter{subsubsection}{0}

\subsubsection{Proof of Theorem~\ref{theorem:charME}}\label{appendix:charproof}

We shall apply Fact~\ref{fact:luenberger} with $V =
\mathcal{L}^1(\rR^d)$, 
\[
F = \{\mathrm{pdf}\, \pi: \mathrm{supp}(\pi) \subseteq S\},
\]
$f(\pi) := -h(\pi)$, ${\bf v}_0 := \pi_0$, and 
\[
g_\gamma(\pi) := \eE_\pi[\phi_\gamma] - u_\gamma
\]
for each $\gamma$ in $\Gamma$. Clearly, $V$ is a vector space and $F$
is a convex subset of $V$.  Since $h(\pi)$ is a concave functional,
$f(\pi)$ is a convex functional on $F$. Also,
$\{g_\gamma(\pi)\}_{\gamma\in \Gamma}$ is a finite collection of
linear (hence convex) functionals on $F$. By assumption, $\pi_0$
belongs to $F$ and for each $\gamma$ in $\Gamma$, $g_\gamma(\pi_0) <
0$. Therefore $F$ is nonempty. The infimum $m_0$ in
Fact~\ref{fact:luenberger} is attained at ${\bf v}^{*} = \pi_{ME}$ and
is equal to $h(\pi_{ME})$ which is finite, that is, $m_0 = h(\pi_{ME})
\in \rR$. Hence $\pi_{ME}\cdot\ln \pi_{ME}$ is absolutely integrable on
$\rR^d$. We have now verified that the conditions of
Fact~\ref{fact:luenberger} are fulfilled and as a consequence, we are
guaranteed the existence of nonnegative reals
$\{\lambda_\gamma\}_{\gamma\in\Gamma}$ so that \eqnref{prob2} and
\eqnref{kkt} hold, that is,
\begin{eqnarray}
-h(\pi_{ME}) = \min_{\pi\in F}\left[-h(\pi) +
  \sum_{\gamma\in\Gamma}\lambda_\gamma
  \left[\eE_\pi[\phi_\gamma]-u_\gamma\right]\right], \eqnlbl{start}
\end{eqnarray}
\begin{eqnarray}
\eqnlbl{active_constraint_cond}
\sum_{\gamma\in\Gamma}\lambda_\gamma \left[\eE_{\pi_{ME}}[\phi_\gamma]
- u_\gamma\right]= 0.
\end{eqnarray}
The last condition is equation \eqnref{qualif} in
Theorem~\ref{theorem:charME}. Consider perturbations around the
minimizer $\pi_{ME}$ of the form 
\[
\pi_{\theta} := \pi_{ME}\cdot(1 + \theta\cdot q)
\]
where $\theta\in[0,1]$,
\begin{equation}
\eqnlbl{perturb} 
q \in \mathcal{L}^{\infty}(\rR^d),\,||q||_{\infty} \leq 1,\ \mathrm{and}\
\eE_{\pi_{ME}}[q] =0.
\end{equation}
It can be verified that $\pi_{\theta}\geq 0,
||\pi_{\theta}||_{\mathcal{L}^1} = 1$, and
$\mathrm{supp}(\pi_{\theta}) \subseteq S$, that is, $\pi_\theta$ is a
pdf with support contained in $S$ for every $\theta\in[0,1]$. This
ensures that the $\theta$--perturbations of $\pi_{ME}$ along $q$ lie
inside $F$. In view of \eqnref{active_constraint_cond}, for all
$\gamma\in\Gamma$ for which $\lambda_\gamma > 0$, we must have
\[
\eE_{\pi_{ME}}[\phi_{\gamma}] = u_\gamma \in \rR
\]
which implies that 
\[
\eE_{\pi_{ME}}\left|\phi_{\gamma}\right| < \infty.
\]
Hence,
\begin{equation}
\eE_{\pi_\theta}\left|\phi_\gamma\right| \leq
(1+\theta)\cdot\eE_{\pi_{ME}}\left|\phi_{\gamma}\right| < \infty
\eqnlbl{finitecond1}
\end{equation}
since $|1+\theta q| \leq 1 + \theta$. Furthermore,
\[
0 \leq (1-\theta) \leq 1 + \theta\cdot q \leq 1 + \theta
\]
implies that for all ${\bf x}\in \rR^d$ and for all $\theta\in [0,1)$,
\[
|\ln(1+\theta\cdot q({\bf x}))| \leq A_\theta :=
\ln\left(\max\left((1+\theta), \frac{1}{(1-\theta)}\right)\right) <
\infty.
\] 
It follows that 
\begin{eqnarray}
|h(\pi_\theta)| &\leq&
(1+\theta)\cdot\eE_{\pi_{ME}}\left|\ln\pi_{ME}\right|\ +\  
\nonumber \\ & & +\ A_\theta <
\infty,\ \forall \theta \in [0,1).\eqnlbl{finitecond2}
\end{eqnarray}
This shows that $\pi_{\theta}\cdot\ln\pi_{\theta}$ is also absolutely
integrable for all $\theta$ in $[0,1)$.  In view of \eqnref{start},
\eqnref{finitecond1}, \eqnref{finitecond2}, and the fact that $\Gamma$
is a finite index set,
\begin{eqnarray*}
\lefteqn{-\infty\ <\ -h(\pi_{ME}) +
\sum_{\gamma\in\Gamma}\lambda_\gamma \left[\eE_{\pi_{ME}}[\phi_\gamma]
- u_\gamma\right]\ \leq} \\ & & \leq\ -h(\pi_{\theta}) +
\sum_{\gamma\in\Gamma}\lambda_\gamma\left[\eE_{\pi_{\theta}}[\phi_\gamma]
-u_\gamma\right]\ <\ +\infty.
\end{eqnarray*}
Collecting terms together and using \eqnref{active_constraint_cond}
we arrive at:
\begin{eqnarray*}
\lefteqn{0\ \leq\
\theta\cdot\eE_{\pi_{ME}}[q\cdot\sum_{\gamma\in\Gamma}\lambda_\gamma\phi_\gamma]\
\ +}\\ && +\ \int_{S}(\pi_{\theta}\cdot\ln \pi_{\theta} - \pi_{ME}\cdot\ln
\pi_{ME})\ <\ +\ \infty.
\end{eqnarray*}
Thus for all $\theta\in (0,1)$ we have
\begin{eqnarray}
\lefteqn{0\ \leq\
\eE_{\pi_{ME}}[q\cdot\sum_{\gamma\in\Gamma}\lambda_\gamma\phi_\gamma]\
\ +}\nonumber\\ && +\ \int_{S}\frac{(\pi_{\theta}\ln \pi_{\theta} -
\pi_{ME}\ln \pi_{ME})}{\theta}\ <\ +\ \infty.  \eqnlbl{midway}
\end{eqnarray}
The function 
\[
\tau_\theta := \frac{(\pi_{\theta}\cdot\ln \pi_{\theta} - \pi_{ME}\cdot\ln
\pi_{ME})}{\theta}
\]
is integrable for each $\theta$ in $(0,1)$ and is nondecreasing in
$\theta$ for each ${\bf x}$ in $S$. Furthermore,
\[
\tau_{0+} := \lim_{\theta\downarrow 0}\tau_\theta =
q\cdot\pi_{ME}\cdot(1+\ln\pi_{ME})
\]
is also integrable. The monotone convergence theorem
\cite[p.~87]{royden} applied to $(\tau_\theta - \tau_{0+})$ shows that
\[
\int_{S}\tau_{0+} = \lim_{\theta\downarrow 0} \int_S \tau_\theta.
\]
From \eqnref{midway} one therefore obtains:
\begin{eqnarray}
\eqnlbl{almostthere}
0 &\leq & \int_{S} q\cdot\pi_{ME}\cdot(1 + \ln\pi_{ME} + \sum_{\gamma\in\Gamma}
\lambda_\gamma\phi_\gamma) < +\infty\nonumber\\
&=& \int_{S} q\cdot\pi_{ME}\cdot(\ln\pi_{ME} + \sum_{\gamma\in\Gamma}
\lambda_\gamma\phi_\gamma) < +\infty,
\end{eqnarray}
since $\eE_{\pi_{ME}}[q] = 0 $ from \eqnref{perturb}. But if
\eqnref{almostthere} holds for $q$ satisfying \eqnref{perturb}, it
also holds for $-q$. We are led to the conclusion that for every $q$
belonging to $\mathcal{L}^{\infty}(\rR^d)$ satisfying $||q||_{\infty}
\leq 1$, whenever $\int_S q\cdot\pi_{ME} = 0$ we must also have 
\[
\int_S q\cdot\pi_{ME}\cdot(\ln \pi_{ME} +
\sum_{\gamma\in\Gamma}\lambda_\gamma\phi_\gamma) = 0.
\]
Thus, for all $q$ belonging to $\mathcal{L}^{\infty}(\rR^d)$, whenever
$\int_S q\cdot\pi_{ME} = 0$ we must also have 
\[
\int_S q\cdot\pi_{ME}\cdot(\ln \pi_{ME} +
\sum_{\gamma\in\Gamma}\lambda_\gamma\phi_\gamma) = 0.
\]

Let $S_{ME}:= \mathrm{supp}(\pi_{ME})$.  Now, ${\bf
1}_{S_{ME}}\cdot\pi_{ME}$, ${\bf 1}_{S_{ME}}\cdot\pi_{ME}\cdot\ln\pi_{ME}$, and
$\{{\bf 1}_{S_{ME}}\cdot\pi_{ME}\cdot\phi_{\gamma}\}_{\gamma\in\Gamma}$ all
belong to $\mathcal{L}^1(\rR^d)$ whose norm--dual
\cite[p.~106]{luenberger} is $\mathcal{L}^{\infty}(\rR^d)$.  If
\[
{\bf 1}_{S_{ME}}\cdot\pi_{ME}\cdot(\ln \pi_{ME} +
\sum_{\gamma\in\Gamma}\lambda_\gamma\phi_\gamma)
\]
does not belong to the one--dimensional closed subspace spanned by
${\bf 1}_{S_{ME}}\cdot\pi_{ME}$, then by the Hahn--Banach theorem
\cite[p.~133]{luenberger}, there exists a bounded linear functional
$q$ on $\mathcal{L}^{1}(\rR^d)$ which vanishes at ${\bf
1}_{S_{ME}}\cdot\pi_{ME}$ but not at 
\[
{\bf 1}_{S_{ME}}\cdot\pi_{ME}\cdot(\ln \pi_{ME} +
\sum_{\gamma\in\Gamma}\lambda_\gamma\phi_\gamma),
\]
that is, there exists a $q$ in $\mathcal{L}^{\infty}(\rR^d)$ such that
$\int_S q\cdot\pi_{ME} = 0$ but
\[
\int_S q\cdot\pi_{ME}\cdot(\ln \pi_{ME} +
\sum_{\gamma\in\Gamma}\lambda_\gamma\phi_\gamma) \neq 0
\]
contradicting the conclusion of the last paragraph.  Hence there
exists a real scalar $\alpha$ such that 
\[
\pi_{ME}\cdot(\ln\pi_{ME} + \sum_{\gamma\in\Gamma}
\lambda_\gamma\phi_\gamma) = -\alpha\pi_{ME}
\]
for all ${\bf x}$ in $S_{ME}$, that is,
\[
\pi_{ME}({\bf x}) = {\bf 1}_{S_{ME}}({\bf x})\cdot\exp\{-\alpha
-\sum_{\gamma\in\Gamma}\lambda_\gamma\phi_\gamma({\bf x})\}.
\]

We shall presently show that for each $\pi$ belongs to $\Omega$ with
$-\infty < h(\pi)$, we have $D(\pi||\pi_{ME}) < +\infty$, that is,
$\pi$, $\pi \ll \pi_{ME}$. In particular, this would mean that
\[
\eE_{\pi}[{\bf 1}_{S\backslash S_{ME}}] = 0
\]
for all $\pi\in\Omega$ with $-\infty < h(\pi)$. To show this, define
\[
\pi_k := \left(1-\frac{1}{k}\right)\pi_{ME} + \frac{1}{k}\pi,
\ k=1,2,\ldots
\]
and note that for each $k$, (i) $\pi_k$ belongs to $\Omega$, (ii) $\pi
\ll \pi_k$ and $\pi_{ME} \ll \pi_k$, and (iii) $\pi_k\longrightarrow
\pi_{ME}$, where the convergence is in the almost everywhere sense and
also under the $\mathcal{L}^1(\rR^d)$ norm. We have
\begin{eqnarray*}
+\infty &>& h(\pi_{ME})\ \geq\ h(\pi_k)\ =\\ &=&
\left(1-\frac{1}{k}\right)h(\pi_{ME}) + \frac{1}{k}h(\pi) +\\ & &
\mbox{} + \left(1 - \frac{1}{k}\right)D(\pi_{ME}||\pi_k) +
\frac{1}{k}D(\pi||\pi_k) \\ &\geq&
\left(1-\frac{1}{k}\right)h(\pi_{ME}) + \frac{1}{k}h(\pi) +
\frac{1}{k}D(\pi||\pi_k),
\end{eqnarray*}
where the first inequality follows from the existence of $\pi_{ME}$
and because $\pi$ belongs to $\Omega$, the second equality is an
identity, and the third inequality follows from the nonnegativity of
cross--entropy (Fact~\ref{fact:divnonneg}). Hence,
\[
h(\pi) + D(\pi||\pi_k) \leq h(\pi_{ME})
\]
for all $k$. Taking limits, noting that $\pi_k$ converges to
$\pi_{ME}$ in norm, and using the lower semi--continuity property of
cross--entropy (Fact~\ref{fact:semicont}) one obtains
\[
h(\pi) + D(\pi||\pi_{ME}) \leq h(\pi_{ME}) < \infty.
\]
Since $h(\pi) > -\infty$, $D(\pi||\pi_{ME}) < \infty$.  The
characterization is now complete.
\endproof

\subsubsection{Proof of Remark~\ref{remarks:basicprops}}\label{appendix:basicprops}
Let ${\bf u}$ map to $\boldsymbol{\lambda}({\bf u})$ and $\pi_{ME}$ be
the maxent pdf in $\Omega({\bf u})$. Define 
\[
\Gamma_0 := \{\gamma\in\Gamma: \lambda_\gamma = 0\}.
\]
Suppose that for all $\gamma$ in $\Gamma_0$, $u'_\gamma \geq u_\gamma$
and for all $\gamma$ in $\Gamma\backslash\Gamma_0$, $u'_\gamma =
u_\gamma$. Let $\pi'_{ME}$ be the maxent pdf in $\Omega({\bf u}')$. We
shall show that $\pi_{ME} = \pi'_{ME}$. Clearly, $\Omega({\bf u})
\subseteq \Omega({\bf u}')$ implies that $h(\pi_{ME}) \leq
h(\pi'_{ME})$. On the other hand, using \eqnref{start} with $\pi =
\pi'_{ME}$ it follows that
\begin{eqnarray*}
h(\pi_{ME}) &\geq& h(\pi'_{ME}) -
\sum_{\gamma\in\Gamma\backslash\Gamma_0}\lambda_\gamma[\eE_{\pi'_{ME}}[
\phi_\gamma] - u_\gamma]\\ &\geq& h(\pi'_{ME})
\end{eqnarray*}
since $\lambda_\gamma > 0$ and 
\[
\eE_{\pi'_{ME}}[ \phi_\gamma] \leq u_\gamma\
\]
for all $\gamma$ in $\Gamma\backslash\Gamma_0$. Thus $h(\pi_{ME}) =
h(\pi'_{ME})$. Since $\pi'_{ME}$ is unique, the result follows.

\subsubsection{Proof of
Theorem~\ref{theorem:converse}}\label{appendix:charconverse}

Let $\alpha$ be the normalization constant for which $\pi_{\exp}$ is a
valid pdf. The condition 
\[
\eE_{\pi}[{\bf 1}_{S\backslash S_{\exp}}({\bf x})] = 0
\]
for all $\pi$ belonging to $\Omega({\bf u})$ for which $-\infty <
h(\pi)$ implies that $\pi \ll \pi_{\exp}$, in particular,
$\mathrm{supp}(\pi) \subseteq \mathrm{supp}(\pi_{\exp})$. Hence,
\[
0\ \leq\ D(\pi||\pi_{\exp})\ =\ \alpha +
\sum_{\gamma\in\Gamma}\lambda_\gamma({\bf
u})\eE_{\pi}[\phi_\gamma({\bf x})] - h(\pi)\ <\ \infty.
\]
This implies that
\begin{eqnarray*}
-\infty\ <\ h(\pi) &\leq& \alpha +
 \sum_{\gamma\in\Gamma}\lambda_\gamma({\bf u})\eE_{\pi}[\phi_\gamma]
 \\ &\leq& \alpha + \sum_{\gamma\in\Gamma}\lambda_\gamma u_\gamma\ <\
 \infty.
\end{eqnarray*}
Since $\pi_{\exp}$ belongs to $\Omega({\bf u})$ and
\[
\sum_{\gamma\in\Gamma}\lambda_\gamma(\eE_{\pi_{\exp}}[\phi_\gamma]-u_\gamma)
= 0,
\]
hence $\eE_{\pi_{\exp}}[\phi_\gamma] = u_\gamma$ for all $\gamma:
\lambda_\gamma > 0$. Thus, for all $\pi$ belonging to $\Omega({\bf
u})$ for which $-\infty < h(\pi)$,
\[
-\infty\ <\ h(\pi)\ \leq\ \alpha + 
\sum_{\gamma\in\Gamma}\lambda_\gamma({\bf
u})\eE_{\pi_{\exp}}[\phi_\gamma]\ =\ h(\pi_{\exp}),
\]
that is,
\[
-\infty\ <\ h(\pi)\ \leq\ h(\pi_{\exp})\ <\ \infty.
\]
Hence, for all $\pi$ in $\Omega({\bf u})$, $h(\pi) \leq h(\pi_{\exp})
< \infty$ and $\pi_{\exp}$ belongs to $\Omega({\bf u})$.
\endproof

\Section{Proof of existence theorems}\label{appendix:existproofs}

\setcounter{subsubsection}{0}

\subsubsection{Proof of Theorem~\ref{theorem:existence}}\label{appendix:existence}

Let $\pi_S({\bf x}) := \frac{1}{|S|}{\bf 1}_S({\bf x})$ (note that
$|S| < \infty$). For all $\pi$ in $F$ we have 
\[
0 \leq D(\pi||\pi_S) = \ln|S| - h(\pi),
\]
that is, 
\[
h(\pi) \leq \ln|S| < \infty. 
\]
Since there exists a pdf $\pi_0$ in $F$ for which $-\infty <
h(\pi_0)$, it follows that 
\[
h(F) := \sup_{\pi\in F}h(\pi) \in \rR,
\]
that is, $h(F)$ is finite.  Let $\{\pi_k\}_{k=1}^{\infty}$ be any
sequence of pdfs in $F$ such that for each $k$, $h(\pi_k) \in \rR$ and
$h(\pi_k) \longrightarrow h(F)$ as $k$ goes to $\infty$. Since $F$ is
$\mathcal{L}^1$--complete, from Fact~\ref{fact:centerofattraction} it
follows that there exists a unique pdf $\pi^*$ in $F$ to which $\pi_k$
converges in norm. Convergence in norm implies convergence in measure
which in turn implies the existence of a subsequence which converges
almost everywhere \cite[Proposition~18, p.~95]{royden}. By passing to
the subsequence we can assume that, without loss of generality,
$\pi_k$ converges to $\pi^*$ almost everywhere (and in norm). The
lower semi--continuity property of cross--entropy
(see~\eqnref{semicont} in Fact~\ref{fact:semicont}) implies that
\begin{eqnarray}
\ln|S| - h(\pi^*) = D(\pi^*||\pi_S) & \leq &
\liminf_{k\rightarrow\infty}D(\pi_k||\pi_S)\nonumber \\ &=&
\ln|S| - \lim_{k\rightarrow\infty}h(\pi_k)\nonumber
\\ &=& \ln|S| - h(F).
\end{eqnarray}
This shows that $h(F) \leq h(\pi^*)$. However, $h(\pi^*) \leq h(F)$
because $\pi^*$ belongs to $F$.  It follows that $h(\pi^*) = h(F)$ and
hence $\pi_{ME} = \pi^*$ is the unique maxent pdf in $F$.  \endproof

\begin{proposition}{(Variational completeness of $\Omega$)}
\label{proposition:closedness}
Let $\Omega({\bf u})$ be as in \eqnref{omegadef}. Let
$\{\phi_\gamma\}_{\gamma\in\Gamma}$ be uniformly bounded from below by
$L\in\rR$. Then $\Omega$ is a convex collection of pdfs which is
complete under the $\mathcal{L}^1(\rR^d)$ norm.
\end{proposition}

\begin{proof}
$\Omega$ is convex because $\eE_\pi[\phi_\gamma]$ is linear in
$\pi$. Let $\{\pi_n\}_{n=1}^{\infty}$ be a Cauchy sequence in $\Omega
\subseteq \mathcal{L}^1(\rR^d)$.  Since ${\mathcal{L}}^{1}({\rR}^{d})$
is complete with respect to the
$||\cdot||_{{\mathcal{L}}^{1}({\rR}^{d})}$ norm \cite[Theorem~6,
p.~125]{royden} and $\{\pi_{n}\}_{n=1}^{\infty}$ is a Cauchy sequence,
there exists $\pi\in\mathcal{L}^1(\rR^d)$ such that $\pi_n$ converges
to $\pi$ under the ${\mathcal{L}}^{1}({\rR}^{d})$ norm. We need to
show that: (i) $\pi \geq 0$ (ii) $\int_{{\rR}^{d}}\pi = 1$, and (iii)
for all $\gamma$ in $\Gamma$, $\eE_\pi[\phi_\gamma] \leq
u_\gamma$. Recall that convergence in
${\mathcal{L}}^{1}({\rR}^{d})$--norm implies convergence in (Lebesgue)
measure which in turn implies the existence of a subsequence
$\pi_{n_{k}}$ converging to $\pi$ almost everywhere in $\rR^{d}$
\cite[Proposition~18, p.~95]{royden}. Since each element of the
subsequence satisfies (i), so does the limit $\pi$. Furthermore,
\begin{eqnarray*}
\left|\int_{{\rR}^{d}}(\pi - 1)\right| &=& \left|\int_{{\rR}^{d}}\pi -
\int_{{\rR}^{d}}\pi_{n}\right| \\ &\leq& \int_{{\rR}^{d}}|\pi-\pi_{n}| \\
&=& ||\pi - \pi_{n}||_{{\mathcal{L}}^{1}({\rR}^{d})} \longrightarrow 0
\end{eqnarray*}
as $n\longrightarrow\infty$ so (ii) holds.  Applying Fatou's lemma
\cite[Theorem~9, p.~86]{royden} to the sequence of non--negative
functions
\[
\pi_n({\bf x})\left[\phi_{\gamma}({\bf x})-L\right]
\]
which converges to
\[
\pi({\bf x})\left[\phi_\gamma({\bf x})-L\right],
\]
gives 
\[
\int_{{\rR}^{d}} \pi\phi_{\gamma} \leq \liminf_{n} \int_{{\rR}^{d}}
\pi_n\phi_{\gamma} \leq u_\gamma.
\]
Hence (iii) also holds, and $\pi$ belongs to $\Omega$.
\end{proof}

\subsubsection{Proof of Corollary~\ref{corollary:existence}}\label{appendix:existbdd}

$\Omega$ is nonempty by assumption, convex by definition, and
$\mathcal{L}^1$--complete by
Proposition~\ref{proposition:closedness}. $S$ has finite volume by
assumption. Since $C$ has nonzero volume and $C\subseteq S$ which has
finite volume, $|C| < \infty$. If 
\[
\pi_{C}({\bf x}) := \frac{1}{|C|}{\bf 1}_C({\bf x})
\]
denotes the distribution that is uniform over the set $C$, it is clear
that $\pi_{C}$ belongs to $\Omega({\bf u})$ and 
\[
h(\pi_{C}) = \ln|C| > -\infty.
\]
Hence by Theorem~\ref{theorem:existence}, 
\[
h(\Omega):= \sup_{\pi\in\Omega}h(\pi)\in \rR, 
\]
that is, $h(\Omega)$ is finite, in fact 
\[
-\infty < \ln|C| \leq h(\Omega) \leq \ln|S| < \infty,
\]
and there exists a unique maxent pdf $\pi_{ME}$ belonging to
$\Omega({\bf u})$ having the exponential form given by
Theorem~\ref{theorem:charME}.  \endproof

\subsubsection{Proof of Theorem~\ref{theorem:existence-2}}\label{appendix:existence-2}

For each $\lambda > 0$, let 
\[
Z_\lambda := ||\exp\{-\lambda \psi \}||_{\mathcal{L}^1(\rR^d)} <
+\infty
\]
and 
\[
\pi_\lambda := (Z_\lambda)^{-1}\exp\{-\lambda \psi \}.
\]
For all $\pi$ in $F$ we have 
\[
0 \leq D(\pi||\pi_\lambda) = \lambda \eE_\pi[\psi] + \ln Z_\lambda -
h(\pi),
\]
that is, 
\[
h(\pi) \leq \lambda \eE_\pi[\psi] + \ln Z_\lambda \leq \lambda u + \ln
Z_\lambda < \infty.
\]
Since there exists a pdf $\pi_0$ in $F$ for which $-\infty <
h(\pi_0)$, it follows that 
\[
h(F) := \sup_{\pi\in F}h(\pi) \in \rR,
\]
that is, $h(F)$ is finite. Since $F$ is $\mathcal{L}^1$--complete,
following the proof of Theorem~\ref{theorem:existence}, there exists a
unique pdf $\pi^*$ in $F$ and a sequence $\pi_k$ in $F$ such that for
each $k$, $h(\pi_k) \in \rR$, $h(\pi_k) \longrightarrow h(F)$ as $k$
goes to $\infty$ and $\pi_k\longrightarrow\pi^*$ both in norm and in
the almost everywhere sense.  The lower semi--continuity property of
cross--entropy (see~\eqnref{semicont} in Fact~\ref{fact:semicont}) and
the moment constraints
\[
\{ \forall \pi \in F, -\infty < L \leq \eE_\pi[\psi] \leq u <
\infty\}
\]
imply that
\begin{eqnarray*}
\lambda \eE_{\pi^*}[\psi] + \ln Z_\lambda - h(\pi^*) &=&
D(\pi^*||\pi_\lambda) \\ &\leq&
\liminf_{k\rightarrow\infty}D(\pi_k||\pi_\lambda) \\ &=&
\liminf_{k\rightarrow\infty}[\lambda \eE_{\pi_k}[\psi] + \ln Z_\lambda \\
&&\mbox{} - h(\pi_k) ] \\ &\leq& \lambda u + \ln Z_\lambda -
\lim_{k\rightarrow\infty} h(\pi_k) \\ &=& \lambda u + \ln Z_\lambda -
h(F).
\end{eqnarray*}
This shows that 
\[
h(F) \leq h(\pi^*) + \lambda \left[u -\eE_{\pi^*}\left[\psi
    \right]\right] \leq h(\pi^*) + \lambda\left(u - L\right)
\]
for all $\lambda >0$. Hence for every $\epsilon >0$ by choosing
$\lambda$ such that $\lambda\left(u - L\right) \leq \epsilon$, we
obtain $h(F) \leq h(\pi^*) + \epsilon$. Thus $h(F) \leq
h(\pi^*)$. However, $h(\pi^*) \leq h(F)$ because $\pi^*$ belongs to
$F$.  It follows that $h(\pi^*) = h(F)$ and hence $\pi_{ME} = \pi^*$
is the unique maxent pdf in $F$.  \endproof

\subsubsection{Proof of Corollary~\ref{corollary:existence-2}}\label{appendix:existgeneral}

$\Omega$ is nonempty by assumption, convex by definition, and
$\mathcal{L}^1$-complete by
Proposition~\ref{proposition:closedness}. Let $C'$ be a subset of $C$
having nonzero but finite volume and 
\[
\pi_{C'}({\bf x}) := \frac{1}{|C'|}{\bf 1}_{C'}({\bf x}).
\]
It is clear that $\pi_{C'}$ belongs to $\Omega({\bf u})$ and
\[
h(\pi_{C'}) = \ln|C'| > -\infty.
\]
Since $\phi_{\gamma_0}$ is uniformly bounded from below by $L\in \rR$,
for all $\pi$ in $\Omega$ we have 
\[
-\infty < L \leq \eE_{\pi}[\phi_{\gamma_0}].
\]
Again, for all $\pi$ in $\Omega$, 
\[
\eE_{\pi}[\phi_{\gamma_0}] \leq u_{\gamma_0} < \infty.
\]
Hence by Theorem~\ref{theorem:existence}, 
\[
h(\Omega):= \sup_{\pi\in\Omega}h(\pi)\in \rR,
\]
that is, $h(\Omega)$ is finite, in fact 
\[
-\infty < \ln|C'| \leq h(\Omega) \leq \inf_{\lambda > 0}
\left[u\lambda + \ln Z_{\lambda}\right] < \infty
\]
where $Z_\lambda$ is as in the proof of
Theorem~\ref{theorem:existence-2}, and there exists a unique maxent
pdf $\pi_{ME}$ belonging to $\Omega({\bf u})$ having the exponential
form given by Theorem~\ref{theorem:charME}.  \endproof

\Section{Proof of
Theorem~\ref{theorem:wellbehavedthm}}\label{appendix:wellbehaved}

\begin{proposition}
\label{proposition:hwelldef}
If pdf $\pi$ belongs to ${\mathcal{L}}^{\infty}(\rR^d)$ then $h(\pi)$
exists and 
\[
-\infty < -\ln ||\pi||_{\mathcal{L}^{\infty}} \leq h(\pi).
\]
If pdf $\pi$ belongs to ${\mathcal{L}}^{2}(\rR^d)$ then $h(\pi)$
exists and 
\[
-\infty < 1-||\pi||_{\mathcal{L}^{2}}^{2} \leq h(\pi).
\]
\end{proposition}

\begin{proof}
Since $||\pi||_{\mathcal{L}^{1}} = 1$ and $\pi$ belongs to
${\mathcal{L}}^{\infty}(\rR^d)$, it follows that $0 <
||\pi||_{\mathcal{L}^{\infty}}$. Also, 
\[
0 \leq \pi({\bf x}) \leq ||\pi||_{\mathcal{L}^{\infty}}
\]
almost everywhere. Thus, 
\[
-\infty < -\pi\ln||\pi||_{\mathcal{L}^{\infty}} \leq -\pi\ln \pi.
\]

Since for all nonnegative $t$, $\ln t \leq t-1$, we have 
\[
\pi({\bf x})-(\pi({\bf x}))^2 \leq -\pi({\bf x})\ln\pi({\bf x})
\]
almost everywhere.  Since $\pi$ belongs to $\mathcal{L}^{2}(\rR^d)$
and $\pi$ is a pdf, the result follows.
\end{proof}

\begin{remark} {\rm
The conditions in the above proposition are not necessary for $h(\pi)$
to exist and be strictly greater than $-\infty$. For example, if
\[
\pi(t) := 1_{(0,1]}(t)\frac{1}{2\sqrt{t}}, 
\]
then $h(\pi) = \ln(\frac{2}{e})$, where $1_{(0,1]}(t)$ is the
characteristic function of the interval $(0,1]$.  The conditions in
proposition~\ref{proposition:hwelldef} do not guarantee that $h(\pi)$
will be finite. For example,
\[
\pi(t) := 1_{[e,\infty)}(t)t^{-1}(\ln t)^{-2}
\]
is both bounded and square integrable which implies that $h(\pi)$
exists but, $h(\pi) = +\infty$ \cite[p.~237]{ashinfothy}.  In the
sequel, we shall derive a general moment condition that ensures that
$h(\pi)$ when it exists, is less that $+\infty$
(Corollary~\ref{corollary:bddhcond}).}
\end{remark}

\begin{proposition}{(Sufficient condition for integrability.)}
\label{proposition:basic}
If $\phi : {\rR}^{d} \rightarrow \rR$ is convex and omnidirectionally
unbounded, then for all strictly positive $a$, 
\[
0 < Z_{\phi}(a) := \int_{{\rR}^{d}} \exp\{-a\phi({\bf
x})\}\,\mathrm{d}{\bf x} < \infty.
\]
In other words, a convex and omni--directionally unbounded function is
stable.
\end{proposition}

\begin{proof}
It is clear that for all real-valued $a$, $0 < Z_{\phi}(a)$.  Since
$\phi({\bf x})$ is unbounded in all directions, there exists a
strictly positive $r$ such that for all ${\bf x}$ satisfying $||{\bf
x}||_{\ell_{1}} > r$ we have $\phi({\bf x}) > \phi({\bf 0})$. Thus,
\[ \inf_{{\bf x}\in{\rR}^{d}}\phi({\bf x}) = \inf_{||{\bf
x}||_{{\ell}_{1}} \leq r}\phi({\bf x}) = \min_{||{\bf
x}||_{{\ell}_{1}}\leq r} \phi({\bf x}) = \phi({\bf x}_{0}) \] for some
${\bf x}_{0}$ in ${\rR}^{d}$ satisfying $||{\bf x}_{0}||_{{\ell}_{1}}
\le r$. The second equality follows because $\phi$ being convex on
${\rR}^{d}$ is continuous, and the closed ball 
\[
\{{\bf x}\in{\rR}^{d}:||{\bf x}||_{{\ell}_{1}} \leq r\} 
\]
is a compact subset of ${\rR}^{d}$. Next, define the function
\[
\psi({\bf x}) := \phi({\bf x} + {\bf x}_{0}) - \phi({\bf
x}_{0}). 
\]
Since ${\bf x}_{0}$ is a global minimizer of $\phi({\bf x})$, it
follows that $\psi({\bf x})$ is non--negative, and attains its global
minimum value of $0$ at the origin.  The function $\psi({\bf x})$ also
inherits the convexity and omni--directional unboundedness properties
of $\phi({\bf x})$. Hence it suffices to demonstrate that for all
strictly positive $a$, $\exp\{-a\psi({\bf x})\}$ is integrable. Since
$\psi({\bf x})$ is non--negative and unbounded in all directions,
there exists a $\rho > 0$ such that for all ${\bf x}$ satisfying
$||{\bf x}||_{\ell_{1}} > \rho$ we have $\psi({\bf x}) > 1$. Now
\[ 
\inf_{||{\bf x}||_{\ell_{1}}=\rho}\psi({\bf x}) = \min_{||{\bf x}
||_{\ell_{1}}=\rho}\psi({\bf x}) = \psi({\bf x}^{*}) \geq 1\ \mathrm{with}
\ ||{\bf x}^{*}||_{{\ell}_{1}}=\rho. 
\] 
The first equality follows from the continuity of $\psi({\bf x})$ and
the compactness of the closed sphere of radius $\rho$ in
${\rR}^{d}$. The last inequality above follows from the way $\rho$ has
been defined. For all ${\bf x}$ in $\rR^d$ having a norm $||{\bf
x}||_{\ell_{1}}$ which is strictly larger than $\rho$, the convexity
of $\psi({\bf x})$ and the definition of ${\bf x}^{*}$ imply that
\[ 
1 \leq  \psi({\bf x}^{*}) \leq  \psi(\frac{\rho{\bf x}}{||{\bf
 x}||_{\ell_{1}}}) \leq \frac{\rho}{||{\bf x}||_{\ell_{1}}}\psi({\bf x})
+ (1-\frac{\rho}{||{\bf x}||_{\ell_{1}}})\psi({\bf 0}).
\]
For all $a > 0$ and for all ${\bf x}$ in $\rR^d$ such that $||{\bf
x}||_{\ell{1}} > \rho$ we have
\[ 
0< a\frac{||{\bf x}||_{\ell{1}}}{\rho} \leq a\psi({\bf x}),
\]
since $\psi({\bf 0}) = 0$. Thus, for all ${\bf x}$ in $\rR^d$ such
that 
\[
||{\bf x}||_{\ell_{1}} > \rho > 0,
\]
\[
\exp\{-a\psi({\bf x})\} \leq \exp\{-\frac{a}{\rho}||{\bf
x}||_{\ell_{1}}\}.
\] 
Finally, since 
\[
||{\bf x}||_{\ell_{1}} := \sum_{i=1}^{d} |{\bf x}(i)|, 
\]
and the exponential function $\exp\{-|t|\},\,t\in\rR$ is integrable
over $\rR$, the result follows.
\end{proof}

\noindent The conditions on $\psi({\bf x})$ in the previous proposition
can be somewhat relaxed as the following corollary demonstrates.

\begin{corollary} \label{corollary:psiprop}
A well--behaved function is stable, that is, if
$\psi:\rR^d\longrightarrow \rR$ is well--behaved, then for all $a >0$,
\[
0 < Z_{\psi}(a) := \int_{\rR^d} e^{-a\psi({\bf x})}\mathrm{d}{\bf x}
< +\infty.
\]
\end{corollary}

\begin{proof}
Since $\psi$ is well--behaved, there exists a convex,
omni--directionally unbounded function $\phi:\rR^d\longrightarrow \rR$
and a nonnegative real number $M$ such that for all ${\bf x}$ in
$\rR^d$ whose norm is strictly larger than $M$ we have $\phi({\bf x})
\leq \psi({\bf x})$ and for all ${\bf x}$ in $\rR^d$ whose norm is no
larger than $M$ we have $\psi({\bf x}) < +\infty$. Now, 
\[
Z_\psi(a) = \int_{\{{\bf x}:||{\bf x}||\leq M\}}e^{-a\psi({\bf x})} +
\int_{\{{\bf x}:||{\bf x}||>M\}}e^{-a\psi({\bf x})}. 
\]
The first term on the right side is bounded since $\psi({\bf x})$ is
bounded over the set 
\[
\{{\bf x}\in\rR^d:||{\bf x}||\leq M\} 
\]
which has finite measure.  Proposition~\ref{proposition:basic}
provides an upper bound for the second term:
\[
\int_{\{{\bf x}:M < ||{\bf x}||\}}e^{-a\psi({\bf x})}\mathrm{d}{\bf x} \leq
\int_{\{{\bf x}:M < ||{\bf x}||\}}e^{-a\phi({\bf x})}\mathrm{d}{\bf x} <
+\infty
\]
and the result follows.
\end{proof}

\begin{proposition}
\label{proposition:bddh}
Let $\pi$ be a pdf for which there exists a convex and
omnidirectionally unbounded function $\phi:\rR^d\rightarrow\rR$ such
that $\phi({\bf x})\leq-\ln\pi({\bf x})$ for all $||{\bf x}||$
sufficiently large. Then $h(\pi)$ exists and $h(\pi) < +\infty$. If
further, $\pi$ belongs to $\mathcal{L}^{\infty}(\rR^d)$ or
$\mathcal{L}^{2}(\rR^d)$ then $h(\pi)$ exists, and $|h(\pi)| <
+\infty$, that is, $-\pi\ln\pi$ belongs to $\mathcal{L}^{1}(\rR^d)$.
\end{proposition}

\begin{proof}
Let 
\[
P:= \{{\bf x}\in\rR^d: 0 \leq \pi({\bf x}) \leq 1\} 
\]
be the set over which $-\pi\ln\pi$ is nonnegative.  From the
assumptions on $\pi$ there exists a strictly positive real number $R$
such that for all $||{\bf x}|| > R$,
\[
0 < \phi({\bf x}) \leq -\ln\pi({\bf x}).  
\]
Define the set 
\[
B := \{{\bf x}\in\rR^d: ||{\bf x}|| > R\}.  
\]
Its complement: $B^c$ is a closed and bounded subset of $\rR^d$ and
has finite volume. Write
\begin{eqnarray}
\eqnlbl{star}
\int_P -\pi\ln\pi = \int_{P\cap B^c}-\pi\ln\pi + \int_{P\cap
B}-\pi\ln\pi.
\end{eqnarray}
We shall show that each integral on the right side of the above
equality is upper bounded by a positive real number. From this it will
follow that $h(\pi)$ exists and $h(\pi) < +\infty$.  Since for all
nonnegative $t$, $\ln t \leq t$, for all ${\bf x}$ in $P$ we have
\[
\begin{array}{c}
0\leq \ln\frac{1}{\pi({\bf x})} \leq \frac{1}{\pi({\bf x})} \\
\Rightarrow 0 \leq -\pi({\bf x})\ln\pi({\bf x}) \leq 1.
\end{array}
\]
Thus the first integral on the right side of \eqnref{star} is upper
bounded by the volume of $P\cap B^c$ which is less than the volume of
the bounded set $B^c$.  Again, since for all nonnegative $t$, $\ln t
\leq \sqrt{t}$, for all ${\bf x}$ in $P$ we have
\[
\begin{array}{c}
0\leq -\ln\pi({\bf x}) \leq \frac{1}{\sqrt{\pi({\bf x})}}\\
\Rightarrow 0 \leq -\pi({\bf x})\ln\pi({\bf x}) \leq \sqrt{\pi({\bf
x})}.
\end{array}
\]
Now for all ${\bf x}$ in $P\cap B$ we have,
\[
\quad 0 \leq -\pi({\bf x})\ln\pi({\bf x}) \leq
\sqrt{\pi({\bf x})} \leq e^{-\frac{\phi({\bf x})}{2}}
\]
where the last inequality follows from the fact that 
\[
\phi({\bf x}) \leq -\ln\pi({\bf x}) 
\]
for all ${\bf x}$ in $B$. We are lead to the following inequalities
\[
0 \leq \int_{P\cap B}-\pi\ln\pi \leq \int_{P\cap B}
e^{-\frac{\phi({\bf x})}{2}} \leq  
||e^{-\frac{\phi({\bf x})}{2}}||_{\mathcal{L}^{1}} < +\infty,
\]
where the last inequality is a consequence of
Corollary~\ref{corollary:psiprop}.  From
Proposition~\ref{proposition:hwelldef} it follows that if further
$\pi$ belongs to ${\mathcal{L}^{\infty}}(\rR^d)$ or to
${\mathcal{L}^{2}}(\rR^d)$, then $h(\pi) > -\infty$ and hence
$|h(\pi)| < +\infty$, that is, $-\pi\ln\pi$ is absolutely
integrable. The proof is complete.
\end{proof}

\begin{corollary}
\label{corollary:genprop}
Let $\psi({\bf x})$ be well--behaved.  If we define 
\[
\pi({\bf x}):= \frac{e^{-\psi({\bf x})}}{Z_{\psi}(1)},
\]
then, $\pi$ is a pdf, $\pi$ belongs to both
$\mathcal{L}^{\infty}(\rR^d)$ and $\mathcal{L}^{2}(\rR^d)$.  Thus
$\pi$ satisfies the conditions and hence the results of
Proposition~\ref{proposition:bddh}, that is, $h(\pi)$ exists and
$|h(\pi)| < +\infty$.
\end{corollary}

\begin{corollary}
\label{corollary:bddhcond}
Let $\phi_{\gamma_0}$ be a well--behaved function. If $\pi$ is any pdf
that satisfies 
\[
\eE_\pi[\phi_{\gamma_0}] \leq u_{\gamma_0} < +\infty 
\]
then $h(\pi)$ exists and 
\[
h(\pi) \leq u_{\gamma_0} + \ln||e^{-\phi_{\gamma_0}}||_{\mathcal{L}^1}
< +\infty.
\]
If further $-\infty < h(\pi)$ then $0\leq D(\pi||r) < +\infty$ where
\[
r({\bf x}) := \frac{e^{-\phi_{\gamma_0}({\bf
x})}}{||e^{-\phi_{\gamma_0}} ||_{\mathcal{L}^1}}
\]
is a pdf.
\end{corollary}

\begin{proof} Corollary~\ref{corollary:psiprop} shows that $r({\bf x})$
is integrable and is hence a valid pdf. It is also clear that
$\mathrm{supp}(r) = \rR^d$ and hence $\pi \ll r$ for each pdf
$\pi$. The inequality $\ln t \leq t-1$ which holds for all nonnegative
$t$ when applied to $t = r({\bf x})\slash \pi({\bf x})$ reveals
that for all ${\bf x}$ belonging to the support--set of the pdf $\pi$,
\begin{eqnarray*}
-\pi({\bf x})\ln\pi({\bf x}) &\leq& r({\bf x}) - \pi({\bf x}) +
\pi({\bf x})\phi_{\gamma_0}({\bf x}) + \\ & &\mbox{} + \pi({\bf x})\ln
||e^{-\phi_{\gamma_0}}||_{\mathcal{L}^1}.
\end{eqnarray*}
Now since $\eE_\pi[\phi_{\gamma_0}] \leq u_{\gamma_0}$ for $\pi$
belonging to $\Omega$, integrating the above inequality over
$\mathrm{supp}(\pi)$ we can conclude that $h(\pi)$ exists and
\[
h(\pi) \leq u_{\gamma_0} + \ln ||e^{-\phi_{\gamma_0}}||_{\mathcal{L}^1}
< +
\infty. 
\]
It is also clear that if $-\infty < h(\pi)$ then
\begin{eqnarray*}
0 &\leq& D(\pi||r) \\ &\leq&
  \ln||e^{-\phi_{\gamma_0}}||_{\mathcal{L}^1} - h(\pi) + \int_{\rR^d}
  \pi\phi_{\gamma_0} \\ &\leq& u_{\gamma_0} + \ln
  ||e^{-\phi_{\gamma_0}}||_{\mathcal{L}^1} - h(\pi) \\ &<& +\infty.
\end{eqnarray*}
\end{proof}

Theorem~\ref{theorem:wellbehavedthm} follows from
Corollary~\ref{corollary:psiprop} and
Corollary~\ref{corollary:bddhcond}.

\begin{biographynophoto}{Prakash~Ishwar} 
(S'98--M'05) received the B. Tech. degree in Electrical Engineering
from the Indian Institute of Technology, Bombay, in 1996, and the
M.S. and Ph.D. degrees in Electrical and Computer Engineering from the
University of Illinois at Urbana-Champaign in 1998 and 2002
respectively.

From August 2002 through December 2004 he was a post-doctoral
researcher in the Electronics Research Laboratory and the department
of Electrical Engineering and Computer Sciences at the University of
California, Berkeley. In January 2005, he joined Boston University
where he is currently an Assistant Professor in the Department of
Electrical and Computer Engineering.  His current research interests
include distributed, decentralized, and collaborative signal
processing, information theory, signal modeling and inference,
multiresolution signal processing, and optimization theory with
applications to sensor networks, multimedia-over-wireless, and
information--security.

Dr.~Ishwar was awarded the 2000 Frederic T. and Edith F. Mavis College
of Engineering Fellowship of the University of Illinois.
\end{biographynophoto}

\begin{biographynophoto}{Pierre~Moulin} 
(F'03) received his D.Sc. from Washington University in St. Louis in
1990. After working for five years as a Research Scientist for Bell
Communications Research in Morristown, New Jersey, he joined the
University of Illinois, where he is currently Professor in the
Department of Electrical and Computer Engineering, Research Professor
in the Coordinated Science Laboratory, faculty member in the Beckman
Institute's Image Formation and Processing Group, and affiliate
professor in the department of Statistics. He is also a member of the
Information Trust Institute. His fields of professional interest are
information theory, image and video processing, statistical signal
processing and modeling, decision theory, information hiding and
authentication, and the application of multiresolution signal
analysis, optimization theory, and fast algorithms to these areas.

In 1996-1998, he served as Associate Editor for the IEEE Transactions
on Information Theory, and in 1999, he was co-chair of the IEEE
Information Theory workshop on Detection, Estimation and
Classification. He was a Guest Editor of the IEEE Transactions on
Information Theory 2000 special issue on Information-Theoretic
Imaging; Guest Editor of the IEEE Transactions on Signal Processing's
2003 special issue on Data Hiding; and member of the IEEE Image and
Multidimensional Signal Processing (IMDSP) Society Technical Committee
(1998-2003). He is currently Area Editor of the IEEE Transactions on
Image Processing and Editor-in-Chief of the upcoming IEEE Transactions
on Information Forensics and Security. He is currently a member of the
Board of Governors of the IEEE Signal processing Society. He has
received a 1997 Career award from the National Science Foundation, and
the IEEE Signal Processing Society 1997 Best Paper award in the IMDSP
area. He is also co-author (with Juan Liu) of a paper that received
the IEEE Signal Processing Society 2002 Young Author Best Paper award
in the IMDSP area. He was selected as 2003 Beckman Associate of UIUC's
Center for Advanced Study, and is a Fellow of IEEE.
\end{biographynophoto}


\begin{thebibliography}{10}
\providecommand{\url}[1]{#1}
\csname url@rmstyle\endcsname
\providecommand{\newblock}{\relax}
\providecommand{\bibinfo}[2]{#2}
\providecommand\BIBentrySTDinterwordspacing{\spaceskip=0pt\relax}
\providecommand\BIBentryALTinterwordstretchfactor{4}
\providecommand\BIBentryALTinterwordspacing{\spaceskip=\fontdimen2\font plus
\BIBentryALTinterwordstretchfactor\fontdimen3\font minus
  \fontdimen4\font\relax}
\providecommand\BIBforeignlanguage[2]{{%
\expandafter\ifx\csname l@#1\endcsname\relax
\typeout{** WARNING: IEEEtran.bst: No hyphenation pattern has been}%
\typeout{** loaded for the language `#1'. Using the pattern for}%
\typeout{** the default language instead.}%
\else
\language=\csname l@#1\endcsname
\fi
#2}}

\bibitem{topsoe79}
F.~Tops\o{e}, ``Information theoretical optimization techniques,''
  \emph{Kybernetika}, vol.~15, pp. 7--17, 1979.

\bibitem{shore-IT80}
{J. Shore and R. Johnson}, ``{Axiomatic derivation of the principle of maximum
  entropy and the principle of minimum cross entropy},'' \emph{IEEE Trans.
  Inform. Theory}, vol. IT--26, pp. 26--37, {Jan.} 1980.

\bibitem{campenhout}
{J. M. Van Campehout and T. M. Cover}, ``{Maximum entropy and conditional
  probability},'' \emph{IEEE Trans. Inform. Theory}, vol. {IT--27}, no.~4, pp.
  483--489, {Jul.} 1981.

\bibitem{jaynes-PrIEEE82}
{E. T. Jaynes}, ``{On the rationale of maximum entropy methods},'' \emph{Proc.
  IEEE}, vol.~70, pp. 939--952, {Sep.} 1982.

\bibitem{csiszar-AnnSt91}
I.~Csisz\'{a}r, ``Why least squares and maximum entropy? {An} axiomatic
  approach to inference for linear inverse problems,'' \emph{Ann. Statist.},
  vol.~19, pp. 2303--2066, 1991.

\bibitem{vogel}
{P. H. A. Vogel}, ``{On the rate distortion function of sources with incomplete
  statistics},'' \emph{IEEE Trans. Inform. Theory}, vol. {IT--38}, no.~1, pp.
  131--135, {Jan.} 1992.

\bibitem{grunwald}
P.~Grunwald, ``The minimum description length principle and reasoning under
  uncertainty,'' Ph.D. dissertation, Universiteit van Amsterdam, Amsterdam, the
  Netherlands, Oct. 1998, available at {\em
  http://www.cwi.nl/$\sim$pdg/thesispage.html}.

\bibitem{borwein}
{J. M. Borwein and M. A. Limber}, ``{On entropy maximization via convex
  programming},'' available at {\em
  http://www.cecm.sfu.ca/$\sim$malimber/IEEE/IEEE.html}.

\bibitem{csiszar-AnnPr75}
I.~Csisz\'{a}r, ``I--divergence geometry of probability distributions and
  minimization problems,'' \emph{Ann. Prob.}, vol.~3, pp. 147--158, 1975.

\bibitem{DowsonW_ME:IT73}
{D. C. Dowson and A. Wragg}, ``{Maximum-entropy distributions having prescribed
  first and second moments},'' \emph{IEEE Trans. Inform. Theory}, vol. {IT-19},
  pp. {689--693}, {Sep.} 1973.

\bibitem{JuppM_AN:ScandJStat83}
{P. E. Jupp and K. V. Mardia}, ``{A note on the maximum-entropy principle},''
  \emph{{Scand. J. Statist.}}, vol.~10, pp. {45--47}, 1983.

\bibitem{csiszar-AnnPr84}
I.~Csisz\'{a}r, ``{Sanov property, generalized $I$-projection and a conditional
  limit theorem},'' \emph{{Ann. Prob.}}, vol.~12, pp. 768--793, {Aug.} 1984.

\bibitem{Kapur_ME93}
{J. N. Kapur}, \emph{{Maximum Entropy Models in Science and
  Engineering}}.\hskip 1em plus 0.5em minus 0.4em\relax New York, NY: John
  Wiley, 1993.

\bibitem{Khudanpur_AM95}
{S. Khudanpur}, ``{A method of ME estimation with relaxed constraints},'' in
  \emph{{Johns Hopkins University Language Modeling Workshop}}, 1995, pp.
  1--17.

\bibitem{royden}
H.~L. Royden, \emph{Real Analysis}.\hskip 1em plus 0.5em minus 0.4em\relax New
  York, NY: Prentice Hall, 1988.

\bibitem{billey}
P.~Billingsley, \emph{Probability and Measure}.\hskip 1em plus 0.5em minus
  0.4em\relax New York, NY: Wiley-Interscience, 1995.

\bibitem{icip99}
{P. Ishwar and P. Moulin}, ``{Multiple--domain image modeling and
  restoration},'' in \emph{Proc. IEEE Int. Conf. on Image Proc.}, vol.~1,
  {Kobe, Japan}, {Oct.} 1999, pp. 362--366.

\bibitem{icassp00}
------, ``{Fundamental equivalences between set--theoretic and maximum--entropy
  methods in multiple--domain image restoration},'' in \emph{Proc. IEEE Int.
  Conf. Acoust., Speech, and Signal Proc.}, vol.~1, {Istanbul, Turkey}, {Jun.}
  2000, pp. 161--164.

\bibitem{icip00}
------, ``{Shift invariant restoration -- an overcomplete maxent MAP
  framework},'' in \emph{Proc. IEEE Int. Conf. on Image Proc.}, vol.~3,
  {Vancouver, Canada}, {Sep.} 2000, pp. 270--272.

\bibitem{my_phd_thesis}
{P. Ishwar}, ``{A Unified Framework for Image Modeling and Estimation using
  Measurement Constraints},'' Ph.D. dissertation, {University of Illinois at
  Urbana-Champaign}, {Jun.} 2002, available at {\em
  http://www.ifp.uiuc.edu/$\sim$moulin/students/ishwar/PhDthesis.ps}.

\bibitem{cover-InfThyJWly}
T.~M. Cover and J.~A. Thomas, \emph{Elements of Information Theory}.\hskip 1em
  plus 0.5em minus 0.4em\relax New York, NY: John Wiley, 1991.

\bibitem{CsiszarM_IP:IT03}
{I. Csisz\'{a}r and F. Mat\'{u}\v{s}}, ``{Information projections revisited},''
  \emph{IEEE Trans. Inform. Theory}, vol. {IT-49}, pp. {1474--1490}, {Jun.}
  2003.

\bibitem{CsiszarM_OI:IT04}
------, ``{On information closures of exponential families: a
  counterexample},'' \emph{IEEE Trans. Inform. Theory}, vol. {IT-50}, pp.
  {922--924}, {May} 2004.

\bibitem{entoptmathprog}
{S. C. Fang, J. R. Rajasekera, and H. S. J. Tsao}, \emph{{Entropy Optimization
  and Mathematical Programming}}.\hskip 1em plus 0.5em minus 0.4em\relax
  Boston: {Kluwer Academic}, 1997.

\bibitem{csiszar-bayes}
I.~Csisz\'{a}r, ``{Maxent, mathematics, and information theory},'' in
  \emph{Maximum Entropy and Bayesian Methods: Proceedings of the 15th
  International Worksop, Santa Fe, New Mexico, USA, 1995}, {K. M. Hanson and R.
  N. Silver}, Ed.\hskip 1em plus 0.5em minus 0.4em\relax {Kluwer Academic},
  1996, pp. 35--50.

\bibitem{bertsekas-NlinPrgAthSc}
D.~P. Bertsekas, \emph{Nonlinear Programming}.\hskip 1em plus 0.5em minus
  0.4em\relax Belmont, Massachusetts: Athena Scientific, 1995.

\bibitem{aliprantis}
C.~D. Aliprantis and O.~Burkinshaw, \emph{Principles of Real Analysis}.\hskip
  1em plus 0.5em minus 0.4em\relax San Diego, CA: Academic Press, 1990.

\bibitem{Kullback:ITandStat}
{S. Kullback}, \emph{{Information Theory and Statistics}}, 2nd~ed.\hskip 1em
  plus 0.5em minus 0.4em\relax {Mineola, NY}: {Dover Publications}, 1968.

\bibitem{pinsker}
{M. S. Pinsker}, \emph{{Information and Information Stability of Random
  Variables and Processes}}.\hskip 1em plus 0.5em minus 0.4em\relax San
  Francisco: {Holden-Day, San Francisco}, 1964.

\bibitem{luenberger}
D.~G. Luenberger, \emph{Optimization by Vector Space Methods}.\hskip 1em plus
  0.5em minus 0.4em\relax New York, NY: John Wiley, 1969.

\bibitem{ashinfothy}
R.~B. Ash, \emph{Information Theory}.\hskip 1em plus 0.5em minus 0.4em\relax
  New York, NY: Dover, 1990.

\end{thebibliography}
\end{document}